\documentclass[12pt,twoside]{article}

\def\Ro{{\bf R}}
\def\Co{{\bf C}}
\def\Io{{\bf I}}

\def\Tr{\,{\rm Tr}\,}

\newtheorem{definition}{Definition}
\newtheorem{postulate}{Postulate}
\newtheorem{example}{Example}

\newcommand{\mathenv}[1]{
{{\small
\par\vskip 0.5cm
\begin{description}\item {$\Rightarrow$\,\,\,} #1\end{description}
\vskip 0.5cm
}}
}

\newcommand{\mathenvtwo}[1]{
{{\small
\par
\begin{description}\item {$\Rightarrow$\,\,\,} #1\end{description}
\vskip 0.5cm
}}
}

\newcommand{\ind}[1]{{\sl #1 }\index{#1}}
\newcommand{\indf}[1]{{\sl #1}\index{#1}}

\begin{document}

\title{Operator formalism of quantum mechanics}

\author{  Jan Naudts\\
          Departement Natuurkunde, Universiteit Antwerpen UIA,\\
          Universiteitsplein 1, 2610 Antwerpen, Belgium\\
          E-mail: Jan.Naudts@ua.ac.be
}

\date{December 25, 2000}

\maketitle
\abstract{
This is the first chapter of a new and unconventional textbook on 
quantum mechanics and quantum field theory. The chapter introduces standard 
quantum mechanics by means of a symmetry principle, without reference to 
classical mechanics. The mathematical foundation of this approach comes 
from a recent paper of Naudts and Kuna on covariance systems.
The standard representation of quantum mechanics is derived.
Next, spin and mass of a quantum particle are explained as labels of
projective representations of the Galilei group.

}

\tableofcontents 
\newpage

\def\Operatorformalism{1}



\section{Classical algebra of functions of position}

\subsection{A no-go statement}

A non-relativistic particle has a position $q$ in euclidean space $\Ro^n$, $n=3$. 
In classical mechanics one assumes that the 
particle has also a velocity $v\in \Ro^n$. Such an assumption cannot be made in 
quantum mechanics: one believes that it is impossible to know at once the 
position of a particle and its velocity. Classically, the knowledge of both 
position and velocity at some given time $t_0$ allows to calculate the orbit 
$t\rightarrow q(t)$ of the particle. In quantum mechanics we must not assume 
that such an orbit exists. Indeed, given the orbit and assuming it to be a 
differentiable function, one would be able to calculate velocities. This would 
mean that both positions and velocities are known, which must not be assumed in 
quantum mechanics. Hence, orbits may not be differentiable. The prospect of 
working with orbits, which in the best case are continuous but not 
differentiable, is not very attractive. Therefore one abandons the concept of a 
classical orbit and tries to proceed in another way.

\subsection{Pure and mixed states}

Let be given any function $f$ of position $q$. If
the particle has position $q$ then the function $f$ can be
evaluated in this position. The resulting number is denoted
$\langle f\rangle$ and is called the
{\sl quantum mechanical expectation value}
\index{expectation value}
in the given state. Obviously, for a particle with position $q$ is
\begin{equation}
\langle f\rangle=f(q)
\label{locpart}
\end{equation}

\mathenv{
In mathematical texts it is common to use the notation $\omega(f)$
instead of $\langle f\rangle$.
Let ${\cal C}_0(\Ro^n)$ denote the $C^*$-algebra of continuous
complex functions of $\Ro^n$ vanishing at infinity.
The position $q$ of the particle determines
a mathematical state $\omega$ of ${\cal C}_0(\Ro^n)$ by
\begin{equation}
\omega(f)=f(q),
\qquad f\in {\cal C}_0(\Ro^n)
\end{equation}
Note that $\omega$ is a pure state.
Its Gelfand-Naimark-Segal (GNS) representation is one-dimensional.
}

The physical state of a one-particle system at a given time $t$ is only partly 
determined by the position $q$ of the particle.
Other information fixing the physical state will 
be discussed in what follows. For this discussion the notion of
reducibility is important. The physical state is {\sl reducible}
\index{reducible state}
if it can be written as a convex linear combination of other physical states.
This means that there exists states, labeled $1$ and $2$, and a
number $\lambda$, strictly positive and strictly less than $1$,
such that
\begin{equation}
\langle f\rangle=\lambda\langle f\rangle_1+(1-\lambda)\langle f\rangle_2
\label{convcomb}
\end{equation}
holds for all functions $f$.
A state which is reducible is also called a {\sl mixed} state.
\index{mixed state}
If it is irreducible then it is a {\sl pure state}.
\index{pure state}
The reason why mixed states are not so relevant is that they
can be decomposed into pure states, and hence, that
they are less fundamental.

In classical mechanics a state which is reducible in the above sense
is also reducible when functions of both position and momentum
are considered. In quantum mechanics a state, which is reducible in the
sense of (\ref{convcomb}), can still be irreducible when the
additional information determining the physical state is
taken into account. In such a situation it corresponds with a
physical state of the particle in which its position
is not well determined.

\mathenv{
In mathematical terms this means that the state $\omega$ of
${\cal C}_0(\Ro^n)$ may be non-pure, but that, because
of the extra information determining the physical state
of the system, the physical state cannot be written as a convex combination
of other physical states. This statement is made more precise in the next
sections.
}

In any case, there exists a (generalized) probability distribution
$\rho$ of $\Ro^n$ such that the following decomposition holds
\begin{equation}
\langle f\rangle=\int_{\Ro^n}\hbox{ d}q\,\rho(q)f(q)
\end{equation}
for all functions $f(q)$.
It is then obvious to suggest that the particle is at position $q$ with probability 
density $\rho(q)$. However, such a probabilistic interpretation is misleading if 
it is not supported by experimental evidence. The theory of quantum mechanics 
can be established without entering these interpretational matters. The price we 
pay for this attitude is that the relation between quantum mechanics and 
experimental physics is not clarified. This is acceptable because anyhow a
profound analysis of quantum experiments is out of the scope of the present
text.

\subsection{Fourier decomposition and convolution \\ algebra}

Later on the Fourier transform of functions $f(q)$
is used. It is given by
\begin{equation}
\tilde f(k)=(2\pi)^{-n}\int_{\Ro^n}\hbox{ d}q\,e^{-ikq} f(q)
\label{ft}
\end{equation}
The inverse transformation is then
\begin{equation}
f(q)=\int_{\Ro^n}\hbox{ d}k\, \tilde f(k) e^{ikq}
\label{backft}
\end{equation}
Note the unusual place of the factor $(2\pi)^{-n}$ and
of the minus sign in the exponential function of 
the forward transformation. Physicists call the function
$q\rightarrow e^{ikq}$ a {\sl plane wave} with {\sl wavevector} $k$.
\index{wavevector}
The formula  (\ref{backft}) is then called the decomposition of the
function $f$ into plane waves.

The convolution between two fourier transformed functions $\tilde f$
and $\tilde g$ is given by
\begin{equation}
(\tilde f\times \tilde g)(k)=\int_{\Ro^n}\hbox{ d}k'\,\tilde f(k')\tilde g(k-k')
\label{convprod}
\end{equation}
The  conjugate $\tilde f^*$ of $\tilde f$ is defined by
\begin{equation}
\tilde f^*(k)=\overline{\tilde f(-k)}
\label{convconj}
\end{equation}
Note that throughout the text a $*$ is used to denote the hermitean
conjugate. In physics literature one uses quite often a $\dagger$ to this
purpose.

\mathenv{
With the operations (\ref{convprod}) and (\ref{convconj})
the linear space ${\cal L}_1(\Ro^n)$ 
of complex integrable functions over $\Ro^n$ becomes an 
involutive Banach algebra. It is known that the inverse Fourier transform $g$
of an integrable function $\tilde g$ belongs to ${\cal C}_0(\Ro^n)$
and that it satisfies $||g||_\infty\le ||\tilde g||_1$
(see e.g.\cite{RW91}, Theorem 7.5).
Hence the inverse Fourier transform is a
continuous morphism of the involutive Banach algebra ${\cal L}_1(\Ro^n)$
into ${\cal C}_0(\Ro^n)$.
}


\section{Symmetries of physical state space}

\subsection{Covariant representations}

Additional information about the physical state of the particle comes from
the requirement that any function of position $f$ is represented as
an operator $\hat f$ on a Hilbert space $\cal H$ and that
symmetry operations on $f$ are represented by unitary transformations
of $\cal H$. Before explaining these points in more detail let us formulate
a few postulates. These postulates should not be taken too absolute.
They are helpful to explain quantum mechanics. As presented here,
they do not form a basis for an axiomatic foundation.

\begin{postulate}[Postulate of reality]
A quantum particle can be observed by measuring
functions of its position.
\end{postulate}

This means that what counts are quantum mechanical expectation values
$\langle f\rangle$ where $f\equiv f(q)$ is a function of the position
$q$ of the particle.

\begin{postulate}[Postulate of covariance]
\label{postcov}
The functions of position must be represented as operators
on a Hilbert space.
Symmetry operations acting on functions of position must be
represented as unitary operators of this Hilbert space.

\end{postulate}

\mathenv{
Also anti-unitary operators should be allowed. An anti-unitary
operator $\theta$ is complex anti-linear:
\begin{equation}
\theta\lambda\psi=\overline\lambda\theta\psi
\end{equation}
and satisfies
\begin{equation}
\langle\theta\psi|\theta\phi\rangle=\langle\phi|\psi\rangle
\end{equation}
} 

\mathenvtwo{
These postulates replace two postulates of conventional quantum mechanics,
one stating that with any measurable quantity corresponds a selfadjoint
operator of a Hilbert space (called an observable),
the other one stating that the result of measuring an observable
returns one of its eigenvalues and leaves the system in the corresponding
eigenstate. The latter statement has not been generally accepted.

} 

An example of a symmetry group of a particle is formed by the shifts in 
position: any position $q$ is shifted to the position $q+a$, with $a$ 
some fixed vector, element of $\Ro^n$. The set of all shifts is the 
additive group $\Ro^n,+$. The action of the shifts can be lifted to the 
algebra of functions of position: Any function $f$ transforms into a 
function $g$ given by $g(q)=f(q-a)$, $q\in\Ro^n$. More generally, we 
require that there is given a group $X$ of relevant symmetries of the 
particle and that this group acts on all functions of the position of 
the particle.

\mathenv{
Technically, the group $X$ will always be assumed to be locally compact.
It acts on the $C^*$-algebra ${\cal C}_0(\Ro^n)$ by means of automorphisms 
$\{\sigma_x:\,x\in X\}$. One has $\sigma_x\sigma_y=\sigma_{xy}$ for all $x,y\in 
X$, and $\sigma_e=1$ ($e$ is the neutral element of $X$). The map $x\in 
X\rightarrow \sigma_x f$ is continuous for all $f\in {\cal C}_0(\Ro^n)$.
}

Consider a representation of functions $f(q)$ as operators $\hat f$ of a 
Hilbert space $\cal H$. In quantum field theory several inequivalent 
representations will be needed, in quantum mechanics only one representation,
{\sl the standard representation of quantum mechanics,} is relevant. It will be 
introduced in the next section.

\mathenv{
In mathematical texts the notation $\pi(f)$ is more common than $\hat f$.
$\pi$  is a *-representation of ${\cal C}_0(\Ro^n)$ as bounded
operators of the Hilbert space $\cal H$. One has
$\pi(fg)=\pi(f)\pi(g)$ and $\pi(f)^*=\pi(\overline f)$ for all
$f,g\in {\cal C}_0(\Ro^n)$.

}

The following definition explains in more detail the contents of
postulate 2.

\begin{definition}
\label{defirrrepr}
The representation $\hat f$ of functions $f$ is $X$-{\sl covariant}
\index{covariant representation}
if for any symmetry transformation $x$ of $X$, which maps a function $f$
into a function $g$,
there exists a unitary operator $U(x)$ such that
$\hat g(q)=U(x)\hat f(q)U(x)^*$ holds.
The covariant representation is said to be {\sl irreducible}
\index{irreducible covariant representation}
if the only operators commuting with all $\hat f$ and all 
$U(x)$ are multiples of the identity operator.
\end{definition}

The transformations $U(x)$ of a covariant representation are
also said to be {\sl canonical} transformations.
The relation between irreducibility of the quantum mechanical
expectation values and irreducibility of covariant representations
is discussed later on.

\mathenv{
The automorphisms $\{\sigma_x,x\in X\}$ are implemented
as unitary operators $\{U(x):\,x\in X\}$ of $\cal H$,
satisfying
\begin{enumerate}
\item {}
 $\pi(\sigma_x f)=U(x)\pi(f)U(x)^*$ for all $x\in X$
and $f\in {\cal C}_0(\Ro^n)$.
\item {}
The map $x\in X\rightarrow U(x)$
is strongly continuous.
\end{enumerate}
}

In general, the unitary operators $U(x)$ do {\sl not} form a representation of 
the group $X$. This means that $U(xy)$ is not necessarily equal to $U(x)U(y)$. 
In this situation one says that $U$ is a {\sl projective} representation 
of $X$.

Let
\begin{equation}
\zeta(x,y)=U(xy)^*U(x)U(y)
\end{equation}
Then, obviously,
\begin{equation}
U(x)U(y)=U(xy)\zeta(x,y)
\label{projective}
\end{equation}
From the definition one can deduce that $\zeta(x,y)$ commutes
with all operators $\hat f$. To show this, assume $x$
transforms $f$ into $g$, and $y$ transforms $g$ into $h$.
Then one has
\begin{eqnarray}
U(yx)\hat fU(yx)^*
&=&\hat h\cr
&=&U(y)\hat g U(y)^*\cr
&=&U(y)U(x)\hat f U(y)^*U(x)^*
\end{eqnarray}
The latter can be written as
\begin{equation}
\hat f\zeta(y,x)=\zeta(y,x)\hat f
\end{equation}
and shows that $\hat f$ commutes with $\zeta(y,x)$.
In the present chapter $\zeta(x,y)$ will always be a complex number
so that it commutes automatically with the operators $\hat f$.

We will always assume that if $e$ is the neutral element of the group $X$
then $U(e)$ is the identity operator.

\mathenv{
With this assumption it is
easy to verify that always $\zeta(e,y)=\zeta(x,e)=1$.
A further property, left as an exercise, is
\begin{equation}
U(y)^*\zeta(z,x)U(y)=\zeta(zx,y)^*\zeta(z,xy)\zeta(x,y)
\end{equation}
This is called the {\sl cocycle} property.
\index{cocycle property}
The factor $\zeta$ appearing in (\ref{projective}) is called
an operator valued right multiplier.
\index{multiplier}

Two cocycles $\zeta$ and $\zeta'$ are {\sl equivalent}
\index{equivalence of cocycles}
if there exist complex numbers $\lambda(x)$ of modulus one
such that
\begin{equation}
\zeta'(x,y)=\lambda(x)\lambda(y)\overline{\lambda(xy)}\zeta(x,y)
\end{equation}
holds for all $x,y$ in $X$.
Note that necessarily $\lambda(e)=1$ holds.
}

\subsection{The postulate of states}

Let be given a representation which is covariant for the
symmetry group $X$.
Any normalized element $\psi$ of the Hilbert space $\cal H$ determines
quantum expectation values $\langle f\rangle$ by
\begin{equation}
\langle f\rangle=\langle \psi|\hat f|\psi\rangle
\label{omegapsi}
\end{equation}
More generally, it determines a set of correlation functions $F(f,x,y)$
by
\begin{equation}
F(f,x,y)=\langle U(y)^*\psi|\hat f|U(x)^*\psi\rangle
\label{chipsi}
\end{equation}
Obviously, one has $\langle\hat f\rangle=F(f,e,e)$
with $e$ the neutral element of $X$.

\mathenv{
We use here the bra-ket notation. It is related to the
mathematical notation for the inner product of the Hilbert
space by
\begin{equation}
\langle\phi|\psi\rangle\equiv(\psi,\phi)
\end{equation}
For any operator $A$ is
\begin{equation}
\langle\phi|A|\psi\rangle\equiv\langle\phi|A\psi\rangle
\end{equation}

}

Let us now postulate that 

\begin{postulate}[Postulate of states]
\label{stateaxiom}
The physical state of a quantum particle is
fully determined by a normalized element $\psi$
(called the {\sl wavefunction})
\index{wavefunction}
in the Hilbert space of an
irreducible $X$-covariant representation of the algebra of functions of
position. Two wavefunctions $\phi$ and $\psi$ determine the same
physical state if there exists an isometry $V$ which
maps $\phi$ onto $\psi$, and commutes with
all operators $\hat f$ and $U(x)$.
\end{postulate}

\mathenv{
An isometry $V$ is a linear operator satisfying
$\langle V\phi|V\psi\rangle=\langle\phi|\psi\rangle$ for all
wavefunctions $\phi,\psi$, i.e.~$V^*V=\Io$.
Because of the assumption that the covariant representation
is irreducible the isometry $V$ is automatically
a unitary operator. Indeed, $VV^*$ is a projection operator
which commutes with all operators $\hat f$ and $U(x)$.
Because the covariant representation is irreducible, there follows that
$VV^*=\Io$.
}

If $\phi$ and $\psi$ are two wavefunctions which determine the same
physical state then they have the same correlation functions.
Indeed, let $V$ be the isometry which maps $\phi$ onto $\psi$ and
which commutes with all operators $\hat f$ and $U(x)$. Then one has
\begin{eqnarray}
F_\psi(f,x,y)
&=&\langle U(y)^*\psi|\hat f|U(x)^*\psi\rangle\cr
&=&\langle U(y)^*V\phi|\hat f|U(x)^*V\phi\rangle\cr
&=&\langle U(y)^*\phi|\hat f|U(x)^*V^*V\phi\rangle\cr
&=&\langle U(y)^*\phi|\hat f|U(x)^*\phi\rangle\cr
&=&F_\phi(f,x,y)
\end{eqnarray}
The converse is also true.
If two wavefunctions have the same correlation functions
then they determine the same physical state. 

\mathenv{
Proof:
Assume that always $F_\phi(f,x,y)=F_\psi(f,x,y)$.
A linear operator $V$ is defined by
\begin{equation}
V\hat f U(x)^*\phi=\hat f U(x)^*\psi
\end{equation}
Indeed, if $\hat f U(x)^*\phi=\hat g U(y)^*\phi$,
then one has for all $h$ and $z$
\begin{eqnarray}
F_\phi(fh,x,z)
&=&\langle U(z)^*\phi|\hat h\hat fU(x)^*\phi\rangle\cr
&=&\langle U(z)^*\phi|\hat h\hat g U(y)^*\phi\cr
&=&F_\phi(gh,y,z)
\end{eqnarray}
But this implies that $F_\psi(fh,x,z)=F_\psi(gh,y,z)$
and hence
\begin{equation}
\langle \hat h^* U(z)^*\psi|\hat fU(x)^*\psi\rangle
=\langle \hat h^* U(z)^*\psi|\hat g U(y)^*\psi\rangle
\end{equation}
Because the covariant representation is irreducible
the Hilbert space is spanned by the elements 
$\hat h^* U(z)^*\psi$. One concludes therefore
that
\begin{equation}
\hat fU(x)^*\psi=\hat g U(y)^*\psi
\end{equation}
This shows that $V$ is well-defined.
That $V$ is an isometry follows from
\begin{eqnarray}
\langle V\hat h^*U(y)^*\phi|V\hat f U(x)^*\phi\rangle
&=&\langle \hat h^*U(y)^*\psi|\hat f U(x)^*\psi\rangle\cr
&=&F_\psi(fh,x,y)\cr
&=&F_\phi(fh,x,y)\cr
&=&\langle \hat h^*U(y)^*\phi|\hat f U(x)^*\phi\rangle
\end{eqnarray}

Clearly, $V$ maps $\phi$ onto $\psi$.
Remains to show that $V$ commutes with all $\hat f$ and
all $U(x)$. Fix a function $f$ and a symmetry $x$ in $X$.
Let $g$ be a function which transforms into $h$ under $x$.
Then one calculates
\begin{eqnarray}
V\hat f U(x)\hat g U(y)\phi
&=&V\hat f \hat h U(x) U(y)\phi\cr
&=&V\hat f \hat h \zeta(x,y) U(xy)\phi\cr
&=&\hat f \hat h \zeta(x,y) U(xy)\psi\cr
&=&\hat f \hat h U(x) U(y)\psi\cr
&=&\hat f U(x)\hat g  U(y)\psi\cr
&=&\hat f U(x) V\hat g U(y)\phi
\end{eqnarray}

Because the functions $\hat g U(y)\phi$ span the Hilbert space
one obtains $V\hat f U(x)=\hat f U(x)V$,
which shows that $V$ commutes with operators $\hat f$
and with operators $U(x)$.
}

Note that the wavefunctions $\phi$ and $e^{i\alpha}\phi$,
with $\alpha$ a real number,
always determine the same physical state.
Indeed, the operator $V=e^{i\alpha}\Io$
is an isometry commuting with all operators of the Hilbert space.
A {\sl ray} of the Hilbert space is a set of all normalized
\index{ray}
elements of $\cal H$ which differ only by a complex phase factor 
from a given normalized wavefunction $\psi$. Traditionally, the 
physical state of a quantum particle is associated with such a 
ray.

Note also that, in principle, the definition of a physical state 
depends on the choice of symmetry group $X$.

\mathenv{
Let $\pi$ and $\pi'$ be two *-representations of ${\cal C}_0(\Ro^n)$ as bounded
operators of the Hilbert spaces $\cal H$ resp.~${\cal H}'$. Assume
$\pi$ and $\pi'$ are $X$-covariant, i.e.~for each $x\in X$
there exist unitary operators $U(x)$ and $U'(x)$ of
of $\cal H$ resp.~${\cal H}'$ implementing the action of
$X$ in ${\cal C}_0(\Ro^n)$. Let $\phi,\phi'$ be normalized elements
of $\cal H$ resp.~${\cal H}'$. Then $\phi$ and $\phi'$ determine
the same physical state if there exists an isomorphism $V$ of the
Hilbert spaces $\cal H$ and ${\cal H}'$ which maps $\phi$ onto $\phi'$,
and which intertwines $\pi$ and $\pi'$ and intertwines $U$ and $U'$.
By the latter is meant that $V\pi(f)=\pi'(f)V$ and $VU(x)=U'(x)V$
for all $f\in {\cal C}_0(\Ro^n)$ and $x\in X$.
}

\subsection{Irreducibility}

It is now time to explain why covariant representations should be 
irreducible (see Def.~(\ref{defirrrepr})). Let us assume that the 
representation is reducible. Then there exists operators which commute 
with all $\hat f$ and all $U(x)$. In particular, there exists a 
projection operator $E$ which commutes with all $\hat f$ and all $U(x)$ 
and which is not zero and not equal to the identity. 
The wavefunctions of the Hilbert space $\cal H$ can now be classified
into three groups. First consider wavefunctions $\psi$
for which $E\psi\not=0$ and $E\psi\not=\psi$. Then
$\lambda=\langle\psi|E\psi\rangle$ satisfies $1>\lambda>0$
and the correlation functions $F(f,x,y)$ can be written as
\begin{eqnarray}
F(f,x,y)&=&\langle U(y)^*\psi|\hat f U(x)^*\psi\rangle\cr
&=&\langle U(y)^*\psi|\hat f U(x)^*E\psi\rangle\cr
& &+\langle U(y)^*\psi|\hat f U(x)^*(1-E)\psi\rangle\cr
&=&\langle U(y)^*E\psi|\hat f U(x)^*E\psi\rangle\cr
& &+\langle U(y)^*(1-E)\psi|\hat f U(x)^*(1-E)\psi\rangle
\end{eqnarray}
To obtain the latter, we have used the property
of projection operators that $E^2=E$ and $(1-E)^2=1-E$,
as well as the assumption that $E$ commutes with all
$\hat f$ and all $U(x)$.
Now let
\begin{equation}
\psi_1=\lambda^{-1/2}E\psi
\qquad\hbox{ and }\quad
\psi_2=(1-\lambda)^{-1/2}(1-E)\psi
\end{equation}
Then $\psi_1$ and $\psi_2$ are wavefunctions which are properly
normalized.
We obtain
\begin{equation}
F(f,x,y)=\lambda F_1(f,x,y)+(1-\lambda)F_2(f,x,y)
\end{equation}
with
\begin{equation}
F_1=\langle U(y)^*\psi_1|\hat f U(x)^*\psi_1\rangle
\end{equation}
and
\begin{equation}
F_2=\langle U(y)^*\psi_2|\hat f U(x)^*\psi_2\rangle
\end{equation}
This shows that the state described by $\psi$ is reducible.
Hence it is not a physical state (see the discussion
at the beginning of the chapter).

There remain the wavefunctions $\psi$ for which $E\psi=0$
or $E\psi=\psi$.
Let $F$ be the orthogonal projection on the subspace ${\cal H}_\psi$
spanned by elements of the form $\hat f U(x)^*\psi$.
This $F$ obviously commutes with all $\hat f$ and
$U(x)$ and satisfies $F\psi=\psi$. Now, the Hilbert space $\cal H$
can be replaced by its subspace ${\cal H}_\psi$. 
In this subspace there are no projection operators $E$ different
from 0 or $\Io$ which commute with all $\hat f$ and with all $U(x)$
and for which $E\psi=0$ or $E\psi=\psi$. Hence,
the requirement that the state described by $\psi$
is irreducible implies that the covariant representation,
reduced to this subspace, is irreducible.


\section{Standard representation of quantum mechanics}

What is known as the standard representation of quantum mechanics
is a representation of the classical algebra of functions
of position which is covariant for the group of shifts in space.
This representation is constructed now.

\subsection{Construction of a covariant representation}

Consider the Hilbert space of functions of position $\psi(q)$
which are quadratically integrable
\begin{equation}
\int_{\Ro^n}\hbox{ d}q\,|\psi(q)|^2 <+\infty
\end{equation}
A function of position $f(q)$ determines an operator $\hat f$
by
\begin{equation}
\hat f\psi(q)=f(q)\psi(q)
\end{equation}
Mapping functions $f(q)$ onto operators $\hat f$ in this way
is a representation of the classical algebra of
functions of position.
\index{representation of the classical algebra}

\mathenv{

The Hilbert space is ${\cal H}={\cal L}_2(\Ro^n,\Co)$.
A *-representation
$\pi$ of the $C^*$-algebra ${\cal C}_0(\Ro^n)$ is defined by
\begin{equation}
\pi(f)\psi(q)=f(q)\psi(q),
\qquad q\in\Ro^n
\end{equation}
for all $f\in{\cal C}_0(\Ro^n)$ and $\psi\in{\cal H}$.

}

The most relevant symmetry group is the group $\Ro^n,+$ of
shifts in position space. Given $a$ in $\Ro^n,+$ a function
$f$ is transformed into the function $g$ given by
\begin{equation}
g(q)=f(q-a)
\label{gdef}
\end{equation}
A representation $U$ of the group of shifts is defined by
\begin{equation}
U(a)\psi(q)=\psi(q-a)
\label{Udef}
\end{equation}
It satisfies
\begin{eqnarray}
U(a)\hat fU(a)^*\psi(q)
&=&\hat fU(a)^*\psi(q-a)\cr
&=&f(q-a)U(a)^*\psi(q-a)\cr
&=&f(q-a)\psi(q)\cr
&=&g(q)\psi(q)\cr
&=&\hat g\psi(q)
\end{eqnarray}
with $g$ given by (\ref{gdef}).
This shows that the representation $\hat f$ of functions $f(q)$
is covariant for the representation $U$ of the group of shifts.

\mathenv{
The covariant representation is irreducible.

Proof:
Let us first prove that the assumption that
\begin{equation}
\langle\phi|\hat f U(a)^*\psi\rangle=0
\end{equation}
holds for all $f$ and $a$ implies that either $\psi=0$
or $\phi=0$.
Because
\begin{equation}
\langle\phi|\hat f U(a)^*|\psi\rangle=
\int_{\Ro^n}\hbox{ d}q\,f(q)\psi(q+a)\overline{\phi(q)}
\end{equation}
one concludes that $\psi(q+a)\overline{\phi(q)}=0$ for 
almost all $q$ and for all $a$.
This implies that either $\phi=0$ or $\psi=0$.

Now let $E$ be a projection operator which commutes with all $\hat f$
and all $U(a)$. Then $E\psi=\psi$ and $E\phi=0$ implies
\begin{eqnarray}
\langle\phi|\hat f U(a)^*\psi\rangle
&=&\langle\phi|\hat f U(a)^*E\psi\rangle\cr
&=&\langle E\phi|\hat f U(a)^*\psi\rangle\cr
&=&0
\end{eqnarray}
so that either $\phi$ or $\psi$ is zero.
This shows that $E$ is either 0 or $\Io$.

} 

\subsection{Physical states}

Expressions (\ref{omegapsi}) and (\ref{chipsi}) can now be made more explicit.
Let $\psi$ be any normalized quadratically integrable function of position.
It determines quantum expectation values of functions of position $f(q)$ by
\begin{eqnarray}
\langle\hat f\rangle&=&\langle\psi|\hat f|\psi\rangle\cr
&=&\int_{\Ro^n} \hbox{ d}q\,f(q)|\psi(q)|^2,
\label{expdef}
\end{eqnarray}
The correlation functions $F(f,a,b)$ are given by
\begin{eqnarray}
F(f,a,b)&=&\langle U(b)^*\psi|\hat f|U(a)^*\psi\rangle\cr
&=&\int_{\Ro^n}\hbox{ d}q\,f(q)\psi(q+a)\overline{\psi(q+b)}
\end{eqnarray}

\begin{example}
\label{exphysstate}
An example of a physical state is given by the following wavefunction
\begin{equation}
\psi(q)=(2\pi\lambda^2)^{-n/4}e^{-|q|^2/4\lambda^2}
\end{equation}
with $\lambda$ any positive number.
One has
\begin{eqnarray}
\langle f\rangle 
&=&(2\pi\lambda^2)^{-n/2}\int_{\Ro^n}\hbox{ d}q\,
f(q)e^{-|q|^2/2\lambda^2}
\end{eqnarray}
and
\begin{eqnarray}
F(f,a,b)
&=&(2\pi\lambda^2)^{-n/2}
\int_{\Ro^n}\hbox{ d}q\,f(q)e^{-|q+a|^2/4\lambda^2}e^{-|q+b|^2/4\lambda^2}\cr
&=&e^{-|a-b|^2)/8\lambda^2}
\int_{\Ro^n}\hbox{ d}k\,\tilde f(k)e^{-ik\cdot(a+b)/2}
e^{-\lambda^2|k|^2/2}
\label{exrescor}
\end{eqnarray}

\end{example}

Each ray of the Hilbert space of the standard representation
determines a distinct physical state.

\mathenv{
Proof

If two wavefunctions $\psi$ and $\phi$ give rise to the same physical
state then they have the same correlation functions. Hence one has
\begin{equation}
\int_{\Ro^n}\hbox{ d}q\,f(q)\phi(q+a)\overline{\phi(q+b)}
=\int_{\Ro^n}\hbox{ d}q\,f(q)\psi(q+a)\overline{\psi(q+b)}
\end{equation}
for all functions $f$ and for all $a$ and $b$.
Decomposition into plane waves gives
\begin{eqnarray}
\int\hbox{ d}k\int\hbox{ d}k'\,e^{ik\cdot a}e^{-ik'\cdot b}\tilde f(k-k')
\bigg[\tilde\phi(k)\overline{\tilde\phi(k')}
-\tilde\psi(k)\overline{\tilde\psi(k')}\bigg]
&=&0\cr
& &
\end{eqnarray}
Because $f$, $a$, and $b$, are arbitrary this implies that
\begin{equation}
\tilde\phi(k)\overline{\tilde\phi(k')}
=\tilde\psi(k)\overline{\tilde\psi(k')}
\end{equation}
for almost all $k$ and $k'$.
The only solution to the latter is that a complex number $\lambda$
exists such that $\phi=\lambda\psi$ holds.

}

\subsection{Characteristic function of a physical state}
\label{ssectcharfun}


Given a wavefunction $\psi$ introduce a functional $\chi$ by
\begin{eqnarray}
\chi(k,q)
&=&e^{-ik\cdot q/2}\int_{\Ro^n}\hbox{ d}q'\,\overline{\psi(q')}e^{ik\cdot q'}\psi(q'-q)\cr
&=&\int_{\Ro^n}\hbox{ d}q'\,e^{ik\cdot q'}\overline{\psi(q'+q/2)}\psi(q'-q/2)
\label{FDef}
\end{eqnarray}
It clearly satisfies $\chi(0,0)=1$ and $\overline{\chi(k,q)}=\chi(-k,-q)$.
For any choice of complex numbers $\lambda_1,\ldots,\lambda_n$ and for
any choice of $k_1,\ldots,k_n$ and $q_1,\ldots,q_n$ in $\Ro^n$ is
\begin{equation}
\sum_{j,j'=1}^n\overline {\lambda_j}\lambda_{j'} e^{i(k_jq_{j'}-k_{j'}q_j)/2}
\chi(k_{j'}-k_j,q_{j'}-q_j)\ge 0 
\label{dposdef}
\end{equation}

With $\chi$ one can calculate quantum expectation values and correlation functions.
Indeed, one has, using decomposition of $f$ into plane waves,
\begin{eqnarray}
F(f,a,b)
&=&\int_{\Ro^n}\hbox{ d}q\,f(q)\psi(q+a)\overline{\psi(q+b)}\cr
&=&\int_{\Ro^n}\hbox{ d}k\,\tilde f(k)
\int_{\Ro^n}\hbox{ d}q\,e^{ik\cdot q}\psi(q+a)\overline{\psi(q+b)}\cr
&=&\int_{\Ro^n}\hbox{ d}k\,\tilde f(k)e^{-ik\cdot (a+b)/2}
\chi(k,b-a)
\label{Ftochar}
\end{eqnarray}
Following Moyal \cite{MJ49} $\chi(k,q)$ is called
the {\sl characteristic function} of the physical state
\index{characteristic function of a physical state}
determined by the wavefunction $\psi$.

In example (\ref{exphysstate}) one has
\begin{equation}
\chi(k,q)=\exp(-\lambda^2|k|^2/2)\exp(-|q|^2/8\lambda^2)
\end{equation}
This follows immediately by comparing (\ref{exrescor}) with (\ref{Ftochar}).

The {\sl Wigner function} $\rho$ (see e.g.~\cite{FG89}, Ch 1 sect 8)
\index{Wigner function}
is related to the characteristic function $\chi$
by a simple Fourier transform 
\begin{eqnarray}
\rho(q,k)
&=&\int_{\Ro^n}\hbox{ d}q'\,
e^{ik\cdot q'}\overline{\psi(q+q'/2)}\psi(q-q'/2)\cr
&=&(2\pi)^{-n}
\int\hbox{ d}k'\int\hbox{ d}q'
\ e^{-ik'\cdot q}e^{ik\cdot q'}\chi(k',q')
\end{eqnarray}
In example (\ref{exphysstate}) one finds
\begin{equation}
\rho(q,k)=2^{n}\exp(-|q|^2/2\lambda^2)\exp(-2\lambda^2|k|^2)
\label{wignerex}
\end{equation}

\mathenv{
The Wigner function $\rho$ has been used as an {\sl ersatz}
for a probability density function. When properly scaled, it is
interpreted as the probability density of a point $(q,k)$
(or $(q,p)$ with $p=\hbar k$) in classical phase space.
Expression (\ref{wignerex}) is indeed a gaussian distribution
in position $q$ with variance $\lambda^2$,
multiplied with a gaussian distribution
in wavevector $k$ with variance $1/4\lambda^2$.
However, in general the Wigner function is {\sl not} a positive function.
This follows immediately because $\chi$ is
in general not positive definite. Indeed, it satisfies
(\ref{dposdef}) which differs from the usual
requirement for positive definiteness by the appearance of an additional
term $\exp(i(k_jq_{j'}-k_{j'}q_j)/2)$.
One concludes that the above interpretation is not correct.

} 



\subsection{Generators of the group of shifts}

The shift operators $U(a)$ form a Lie group with generators $K_j$,
$j=1,2,\cdots,n$:
\begin{equation}
U(a)=\exp\left(i\sum_{j=1}^na_j K_j\right)
\label{shiftgen}
\end{equation}

\mathenv{
The map $\lambda\in\Co\rightarrow U(\lambda a)$ is strongly continuous.
Take $a$ equal to one of the
the basis vectors $e_j$, $j=1,2,\ldots,n$
of cartesian coordinates in $\Ro_n$.
Then by Stone's theorem there exists a unique self-adjoint operator
$K_j$ such that
\begin{equation}
U(\lambda e_j)=e^{i\lambda K_j}
\end{equation}
for all real $\lambda$.
Because the group $\Ro^n,+$ is abelian (\ref{shiftgen}) follows.
}

Expansion of (\ref{shiftgen}) for small values of $a$ gives
\begin{equation}
U(a)=\Io+i\sum_{j=1}^na_j K_j+\cdots
\end{equation}
with $\Io$ the identity operator.
On the other hand, expansion of (\ref{Udef}) gives
\begin{equation}
U(a)\psi(q)=\psi(q)-\sum_{j=1}^na_j
\frac{\partial\,}{\partial q_j}\psi(q)
+\cdots
\end{equation}
Comparison of both expressions gives
\begin{equation}
K_j\psi(q)=-i\frac{\partial\,}{\partial q_j}\psi(q)
\label{kexplicit}
\end{equation}

\mathenv{
The operator $K_j$ is an unbounded self-adjoint operator
with a domain which should be specified  carefully.
The previous expression is not valid for all
wavefunctions $\psi$ of the Hilbert space, but
is true if the partial derivative  of $\psi(q)$ 
is again a square integrable function.
}

Let $Q_j$, $j=1,2,\cdots,n$, denote the multiplication operators.
They are defined by
$Q_j\psi(q)=q_j\psi(q)$ for all wavefunctions $\psi$ for
which the function $q_j\psi(q)$ is square integrable.
For further use let us calculate
\begin{eqnarray}
\hat f\psi(q)
&=&f(q)\psi(q)\cr
&=&\int_{\Ro^n}\hbox{ d}k\,\tilde f(k)e^{ikq}\psi(q)\cr
&=&\int_{\Ro^n}\hbox{ d}k\,\tilde f(k)e^{ikQ}\psi(q)
\end{eqnarray}
Hence one obtains
\begin{equation}
\hat f=\int_{\Ro^n}\hbox{ d}k\,\tilde f(k)e^{ikQ}
\end{equation}

It is now straightforward to derive the canonical commutation relations (CCR).
From definition (\ref {Udef}) follows
\begin{eqnarray}
U(a)Q\psi(q)
&=&Q\psi(q-a)\cr
&=&(q-a)\psi(q-a)\cr
&=&(q-a)U(a)\psi(q)\cr
&=&(Q-a)U(a)\psi(q)
\end{eqnarray}
Hence, on the domain of definition of $Q$ holds
\begin{equation}
U(a)Q=(Q-a)U(a)
\end{equation}
i.e.
\begin{equation}
U(a)QU(a)^*=e^{-iaK}Qe^{iaK}=Q-a
\label{qshift}
\end{equation}
This implies
\begin{equation}
i[K,Q]=\Io
\label{kqccr}
\end{equation}
with $\Io$ the identity operator.
To see this, make an expansion of (\ref {qshift})
for small values of $a$.

\mathenv{
Traditionally, one defines the momentum operator $P$ in
such a way that $P=\hbar K$ holds
with $\hbar$ Planck's constant. Then one obtains
the CCR in their well-known form
\begin{equation}
[P,Q]={\hbar\over i}\Io
\label{ccr}
\end{equation}

}

Usually one associates Planck's constant $\hbar$, which is a 
very small number, with a quantum deformation of classical 
mechanics. In particular the canonical commutation relations 
(\ref{ccr}) are interpreted in this way. This is a miss-
conception. Planck's constant enters into the formula if one 
tries to replace the classical notion of momentum by a quantum 
mechanical one using the formula $P=\hbar K$ which comes from 
quantum electrodynamics. As shown above, the canonical 
commutation relations can be written as $i[K,Q]=\Io$. No 
constant of Planck is involved and the result of the commutation 
is not a small number.

\subsection{Von Neumann uniqueness theorem}

Von Neumann's theorem states that the standard representation of 
quantum mechanics is unique up to unitary equivalence. Up to now 
we considered functions of position as basic observable 
quantities, and shifts in position as basic symmetry elements. 
Then the only covariant representation which is irreducible is 
the standard representation, up to unitary equivalence. The 
latter means that if another representation exists in some
Hilbert space ${\cal H}'$ then there exists
an isometry $V$ of the Hilbert space ${\cal H}$ of the standard representation
into ${\cal H}'$ which maps operators $(\hat f)_{\cal H}$ onto operators
$(\hat f)_{{\cal H}'}$,
and operators $U(x)$ onto operators $U(x)'$.

In the sections that follow the group of symmetries $X$ will be extended. 
This will change the notion of irreducibility. As a consequence 
the unicity will get lost and other representations like the 
spinor representation will come into play.

\mathenv{
Proof:
The proof is similar to the proof, in the section about the postulate of 
states, that, if two wavefunctions determine the same correlation 
functions, then they determine the same physical state.
The trick is to find wavefunctions $\psi$ in $\cal H$ and
$\psi'$ in ${\cal H}'$ which determine the same correlation functions.
Then, by a slight generalization of the above mentioned proof,
one concludes that an isometry exists which intertwines the
two representations.

Let
\begin{equation}
g(q)=2^{n/2}e^{-|q|^2}
\end{equation}
and
\begin{equation}
G_a(q)=g(q-a/2)
\end{equation}
An operator $E$ is defined by
\begin{equation}
E=\int_{{\Ro}^n}\hbox{ d}a\,e^{-|a|^2/4}\hat G_a U(a)^*
\end{equation}
Using the identity
\begin{equation}
\int_{\Ro^n}\hbox{ d}k\,
e^{-|k|^2/2}e^{iak}=(2\pi)^{n/2}e^{-|a|^2/2}
\label{gaussident}
\end{equation}
one can write $f$ as
\begin{equation}
f(q)=(2\pi)^{-n/2}\int_{\Ro^n}\hbox{ d}k\,
e^{-|k|^2/4}e^{iqk}
\end{equation}
Hence one can write $E$ in the following way
\begin{eqnarray}
E&=&\int_{{\Ro}^n}\hbox{ d}a\,e^{-|a|^2/4}
(2\pi)^{-n/2}\int_{\Ro^n}\hbox{ d}k\,
e^{-|k|^2/4}e^{ik(Q-a/2)}
e^{-iaK}\cr
&=&(2\pi)^{-n/2}\int_{{\Ro}^n}\hbox{ d}a\,e^{-|a|^2/4}
\int_{\Ro^n}\hbox{ d}k\,e^{-|k|^2/4}e^{i(kQ-aK)}
\label{vNproj}
\end{eqnarray}
To obtain the latter we have used the well-known result
\begin{equation}
e^{i(A+B)}=e^{iA}e^{iB}e^{(1/2)[A,B]}
\label{BCH}
\end{equation}
which is valid for any pair of operators $A$ and $B$
satisfying $[B,[B,A]]=0$
and is a special case of the Baker-Campbell-Hausdorff
formula.
Expression (\ref{vNproj}) shows that $E$ is a symmetric
operator.

Now calculate, using again (\ref{BCH}) and (\ref{gaussident}),
\begin{eqnarray}
E^2&=&(2\pi)^{-n}\int_{{\Ro}^n}\hbox{ d}a\,e^{-|a|^2/4}
\int_{\Ro^n}\hbox{ d}k\,e^{-|k|^2/4}e^{i(kQ-aK)}\cr
& &\times \int_{\Ro^n}\hbox{ d}a'\,e^{-|a'|^2/4}
\int_{\Ro^n}\hbox{ d}k'\,e^{-|k'|^2/4}e^{i(k'Q-a'K)}\cr
&=&(2\pi)^{-n}\int_{{\Ro}^n}\hbox{ d}a\,e^{-|a|^2/4}
\int_{\Ro^n}\hbox{ d}k\,e^{-|k|^2/4}\cr
& &\times \int_{\Ro^n}\hbox{ d}a'\,e^{-|a'|^2/4}
\int_{\Ro^n}\hbox{ d}k'\cr
& &\times e^{-|k'|^2/4}e^{i((k+k')Q-(a+a')K)}e^{i(ka'-k'a)/2}\cr
&=&(2\pi)^{-n}\int_{{\Ro}^n}\hbox{ d}a\,
\int_{\Ro^n}\hbox{ d}k\,\int_{\Ro^n}\hbox{ d}a'\int_{\Ro^n}\hbox{ d}k'\,\cr
& &\times e^{-|a-a'|^2/4}e^{-|k-k'|^2/4}e^{-|a'|^2/4}\cr
& &\times
e^{-|k'|^2/4}e^{i(kQ-aK)}e^{i(ka'-k'a)/2}\cr
&=&\int_{{\Ro}^n}\hbox{ d}a\,e^{-|a|^2/4}
\int_{\Ro^n}\hbox{ d}k\,e^{-|k|^2/4}e^{i(kQ-aK)}\cr
&=&E
\end{eqnarray}
This shows that $E$ is a projection operator.

The previous calculation can be generalized to show that
for any function $f$ and any shift $a$ one has
\begin{equation}
E\hat f U(a)E=\lambda(f,a)E
\end{equation}
with
\begin{equation}
\lambda(f,a)=e^{-|a|^2/4}\int_{\Ro^n}\hbox{ d}k\,\tilde f(k)e^{-|k|^2/4}
\end{equation}

Now, the proof of the theorem is almost finished.
Let $\psi$ be a wavefunction in the range of $E$,
i.e.~satisfying $E\psi=\psi$. Then one has
\begin{eqnarray}
F(f,a,b)&=&\langle U(b)^*\psi|\hat f U(a)^*\psi\rangle\cr
&=&\langle\psi|\hat f_{-b}U(b)U(a)^*\psi\rangle\cr
&=&\xi(b,-a)\langle\psi|\hat f_{-b}U(b-a)\psi\rangle\cr
&=&\xi(b,-a)\langle\psi|E\hat f_{-b}U(b-a)E\psi\rangle\cr
&=&\xi(b,-a)\lambda(f_{-b},b-a)\langle\psi|E\psi\rangle\cr
&=&\xi(b,-a)\lambda(f_{-b},b-a)
\end{eqnarray}
This shows that in any covariant representation there exists
a wavefunction $\psi$ which determines a state with
correlation functions given by 
\begin{equation}
F(f,a,b)=\xi(b,-a)\lambda(f_{-b},b-a)
\end{equation}
As discused before, this implies uniqueness
of the standard representation of quantum mechanics,
up to unitary equivalence.

} 


\section{Rotation symmetry}

\label{rotsymsection}

In this section rotation symmetry is studied with the
intention to clarify the mechanical spin of a quantum particle.

\subsection{Group of shifts and rotations}

Besides the group of shifts one can also consider
rotations in space as basic symmetry operations.
The group law is given by
\begin{equation}
(b,M)(a,\Lambda)=(b+M a, M\Lambda)
\end{equation}
As before, $a$ and $b$ are vectors by which the particle is shifted,
$\Lambda$ and $M$ are elements of the rotation group SO$(n)$.

\mathenv{
Note that the group $X$ is now a semidirect product
of $\Ro^n,+$ and SO$(n)$, not a direct product.
One has in particular
\begin{equation}
(0,\Lambda)(a,{\bf 1})=(\Lambda a,\Lambda)
\end{equation}
while
\begin{equation}
(a,{\bf 1})(0,\Lambda)=(a,\Lambda)
\end{equation}

} 

We continue working with the standard representation of quantum mechanics.
A unitary representation $U$ of this group is defined by
\begin{equation}
U(a,\Lambda)\psi(q)=\psi(\Lambda^{-1}(q-a))
\end{equation}

\mathenv{
Indeed, one verifies that $U$ is unitary
\begin{eqnarray}
\langle U(a,\Lambda)\phi|U(a,\Lambda)\psi\rangle
&=&\int_{\Ro^n}\hbox{ d}q\,
U(a,\Lambda)\psi(q)\overline{U(a,\Lambda)\phi(q)}\cr
&=&\int_{\Ro^n}\hbox{ d}q\,
\psi(\Lambda^{-1}(q-a))\overline{\phi(\Lambda^{-1}(q-a))}\cr
&=&\int_{\Ro^n}\hbox{ d}q\,
\psi(\Lambda^{-1}q)\overline{\phi(\Lambda^{-1}q)}\cr
&=&\int_{\Ro^n}\hbox{ d}q\,
\psi(q)\overline{\phi(q)}\cr
&=&\langle\phi|\psi\rangle
\end{eqnarray}
That $U$ is a representation follows from
\begin{eqnarray}
U(b,M)U(a,\Lambda)\psi(q)
&=&U(a,\Lambda)\psi(M^{-1}(q-b))\cr
&=&\psi(\Lambda^{-1}(M^{-1}(q-b)-a))\cr
&=&\psi((M\Lambda)^{-1}(q-b-Ma)\cr
&=&U(b+Ma,M\Lambda)\psi(q)
\end{eqnarray}

} 

Next one verifies that for any wavefunction $\psi$ one has
\begin{eqnarray}
U(a,\Lambda)\hat f U(a,\Lambda)^*\psi (q)
&=&\hat f U(a,\Lambda)^*\psi (\Lambda^{-1}(q-a))\cr
&=&f(\Lambda^{-1}(q-a)) U(a,\Lambda)^*\psi (\Lambda^{-1}(q-a))\cr
&=&f(\Lambda^{-1}(q-a))\psi(q)\cr
&=&\hat g\psi(q)
\end{eqnarray}
with $g(q)=f(\Lambda^{-1}(q-a))$.

One concludes that the standard representation of quantum mechanics is
also covariant for the enlarged group of rotations and shifts all together.
The correlation functions are now given by
\begin{eqnarray}
F(f,(a,\Lambda),(b,M))
&=&\langle U(b,M)^*\psi|\hat f| U(a,\Lambda)^*\psi\rangle\cr
&=&\int_{\Ro^n}\hbox{ d}q\,f(q)\psi(\Lambda q+a)
\overline{\psi(Mq+b))}
\end{eqnarray}
It is left as an exercise to show that $F(f,(a,\Lambda),(b,M))$ can be
expressed in terms of correlation functions $F(f,a,b)$ which involve
only shifts in space $a$ and $b$.

With the basic symmetry group consisting of shifts and rotations the 
standard representation of quantum mechanics is not any longer unique, 
i.e.~ von Neumann's uniqueness theorem does not hold in this case. 
Another representation exists which is known as the spinor 
representation. It is discussed in the next section.

\subsection{Path depending representations}

A particle like an electron has an intrinsic {\sl spin}.
\index{spin of the electron}
This suggests that the particle spins around an axis
which is hard to believe because as far as is known the
electron is a point particle without internal structure.
The mathematical characterization of the spin of the electron
involves a so-called {\sl spinor representation}
\index{spinor representation}
of the rotation group SO$(3)$ instead of the standard representation
used in section \ref{rotsymsection}.

The spinor representation is a projective
representation. In the physics literature one often speaks of a {\sl two-valued
representation},
\index{two-valued representation}
althoug this is a slightly different concept.
The latter is best described as a {\sl path depending representation}
\index{path depending representation} because the
representation of an element of the rotation group depends
on the path by which it is reached, starting from the identity
transformation. 
The existence of such path depending representations is a
consequence of the fact that
the rotation group SO$(3)$ is {\sl double connected},
\index{double connected}
which means that all closed paths
in the group belong to one of two classes. In the first class
all closed paths starting and ending with the
identity transformation can be reduced by continuous deformation to
the path of length zero. In the other class such a reduction is not possible.
For example, any path winding by $2\pi$ around an arbitrary axis
belongs to the latter class. On the other hand one can show that
the path winding by $4\pi$ around any axis belongs to the former class.

\subsection{The universal covering group}

The \ind{universal covering group} of SO$(3)$ is SU$(2)$, the group
of unitary two-by-two matrices with determinant 1.
By definition this means that SU$(2)$ is simply connected and that
there exists a map from SU$(2)$ to SO$(3)$ which is continuous
and which is locally an isomorphism of groups. This map is not unique.
The conventional choice is constructed as follows.

Let $\sigma_1,\sigma_2,\sigma_3$ denote the Pauli matrices
\begin{equation}
\sigma_1=\left(\matrix{0 &1\cr 1 &0}\right)
\quad
\sigma_2=\left(\matrix{0 &-i\cr i &0}\right)
\quad
\sigma_3=\left(\matrix{1 &0\cr 0 &-1}\right)
\end{equation}
They satisfy $\sigma_j^2=\Io$ ($j=1,2,3$).
With the Pauli matrices one associates with each position
$q$ in $\Ro^3$ a 2-by-2 matrix $M(q)$ by
\begin{equation}
M(q)=\sum_{j=1}^3q_j\sigma_j
\end{equation}
Matrices of this type have a vanishing trace $\Tr M(q)=0$.
Now let $u$ be an element of SU(2).
The transformed matrix $uM(q)u^*$ has again a vanishing trace.
Hence it can be written into the form $M(q')$ with $q'$ another vector of $\Ro^3$.
Now, the map from $q$ to $q'$ is a rotation.
Let $\Xi(u)$ denote this rotation: $q'=\Xi(u)q$.

\mathenv{

To see this we show that the scalar
product between two vectors $q$ and $q''$ does not change when they transform to
$q'$ resp.~$q'''$:
\begin{eqnarray}
q'\cdot q'''
&=&\sum_{j=1}^3 q'_j q'''_j\cr
&=&\frac{1}{2}\Tr M(q')M(q''')\cr
&=&\frac{1}{2}\Tr uM(q)u^*uM(q'')u^*\cr
&=&\frac{1}{2}\Tr M(q)M(q'')\cr
&=&q\cdot q''
\end{eqnarray}

It is clear that $\Xi(\Io)=\Io$ and that
\begin{equation}
\Xi(u)\Xi(v)=\Xi(uv)
\end{equation}
Hence the map $\Xi$ is a homomorphism of SU$(2)$ in SO$(3)$.

Assume now that $u$ in SU$(2)$ exists for which $\Xi(u)$
is inversion of $\Ro^3$. Then 
\begin{equation}
uM(q)u^*=-M(q)
\label{inverse}
\end{equation}
holds for all $q$. This implies that $u$ commutes with all
2-by-2 matrices. Now, the only matrices commuting
with all 2-by-2 matrices are multiples $\lambda\Io$
of the identity matrix $\Io$. But $u=\lambda\Io$
in combination with (\ref{inverse}) implies $|\lambda|^2=-1$
which has no solution. One concludes that inversion is
not in the range of $\Xi$ so that all $\Xi(u)$
are proper rotations.

}

The matrix $-\Io$ can also be represented as $U(x)$ with $x$ any vector
of length $2\pi$.  With a vector $x\not=0$ one can associate a rotation with
angle $|x|$ and rotation axis $x/|x|$. With $x=0$ one associates the identity.
Clearly, this realizes a map from SU$(2)$ to SO$(3)$. However, it is not one-to-one.
Rotations with angles $|x|\le\pi$ suffice already to obtain all of SO$(3)$.

Consider now a path in SO$(3)$ from the identity to some arbitrary 
rotation $\Lambda$. With $\Lambda$ correspond two elements 
$U(x)$ of SU$(2)$. However, if one starts with mapping the 
identity of SO$(3)$ onto the identity of SU$(2)$ and follows the 
path linking the identity with $\Lambda$ then there is a unique 
corresponding path in SU$(2)$. It ends at one of the elements 
$U(x)$ representing $\Lambda$. Hence the path determines 
uniquely which element of SU$(2)$ corresponds with $\Lambda$.
In particular, consider now a closed path starting and ending at
the identity of SO$(3)$. Then the corresponding path in SU$(2)$
is either closed, returning to $\Io$, or open,
ending at $-\Io$.

\subsection {A projective representation of SO$(3)$}

It is clear that $\Xi(-u)=\Xi(u)$. Hence the inverse of the map $\Xi$ from SU$(2)$
to SO$(3)$ is not unique. Let $v(\Lambda)$ be such an inverse.
This means that $\Xi(v(\Lambda))=\Lambda$ for all rotations $\Lambda$.
We can chose this inverse so that small rotations $\Lambda$ are identified
with elements $u$ of SU(2) which differ not too much
from the identity matrix. The latter means that $v(\Xi(u))=u$
for $u$ in the neighborhood of the identity matrix $\Io$.

The map $v$ is a projective representation of SO$(3)$ as unitary 2-by-2
matrices. The multipier $\xi(\Lambda,\Lambda')$ is defined by
\begin{equation}
\xi(\Lambda,\Lambda')=v(\Lambda)v(\Lambda')v(\Lambda\Lambda')^*
\end{equation}
Now calculate
\begin{eqnarray}
\Xi(\xi(\Lambda,\Lambda'))
&=&\Xi(v(\Lambda)v(\Lambda')v(\Lambda\Lambda')^*)\cr
&=&\Xi(v(\Lambda))\Xi(v(\Lambda'))\Xi(v(\Lambda\Lambda')^*)\cr
&=&\Lambda\Lambda'(\Lambda\Lambda')^{-1}\cr
&=&\Io
\end{eqnarray}
But the only $u$ in SU(2) for which $\Xi(u)=\Io$ are
$u=\pm \Io$. One concludes that
\begin{equation}
\xi(\Lambda,\Lambda')=\pm 1
\end{equation}

\mathenv{
$\Xi(u)=\Io$ means $M(q)=uM(q)u^*$ for all $q$.
The only elements $u$ of SU(2)
which commute with all 2-by-2 matrices are $u=\pm \Io$.
Hence, $\Xi(u)=\Io$ implies $u=\pm \Io$.
}

\subsection{A spinor representation of SO$(3)$}

A {\sl spinor} is a pair of quadratically integrable functions
\index{spinor}
$|\psi_0,\psi_1\rangle$ normalized by
\begin{equation}
||\psi_0||^2+||\psi_1||^2=1
\label{normspinor}
\end{equation}

\mathenv{

Nonrelativistic spinors belong to the Hilbert space
\begin{equation}
{\cal L}_2(\Ro^n,\Co)\oplus{\cal L}_2(\Ro^n,\Co)
\end{equation}

}

A representation of the classical algebra of functions of position
is obtained from the standard representation in the obvious way
\begin{equation}
\hat f |\psi_0,\psi_1\rangle=
|\hat f\psi_0,\hat f\psi_1\rangle
\end{equation}
In a similar way one obtains a representation of the group of shifts
\begin{equation}
U(a)|\psi_0,\psi_1\rangle=
|U(a)\psi_0, U(a)\psi_1\rangle
\end{equation}
Now let $\Lambda$ be a rotation in SO$(3)$ and $v(\Lambda)$ the
corresponding element of SU$(2)$ then the action $U(\Lambda)$
on the spinor $|\psi_0,\psi_1\rangle$ is defined by
\begin{eqnarray}
U(\Lambda)|\psi_0,\psi_1\rangle
&=&|v(\Lambda)_{0,0}U(\Lambda)\psi_0+v(\Lambda)_{0,1}U(\Lambda)\psi_1,\cr
& &
v(\Lambda)_{1,0}U(\Lambda)\psi_0+v(\Lambda)_{1,1}U(\Lambda)\psi_1\rangle
\end{eqnarray}
where $U(\Lambda)$ is the representation of the rotation group in
the standard representation. In other words, the
projective representation $v(\Lambda)$ of the previous subsection
is combined with the representation $U(\Lambda)$ of the
standard representation of quantum mechanics.

The above definition gives a projective representation of SO$(3)$.
It is obvious that
\begin{equation}
U(\Lambda)U(\Lambda')=\xi(\Lambda,\Lambda')U(\Lambda\Lambda')
\end{equation}
with phase factor $\xi(\Lambda,\Lambda')$ equal to $\pm 1$.
If the rotations $\Lambda$ and $\Lambda'$ are small enough then
$\xi(\Lambda,\Lambda')=1$ holds.

For further usage note that the conjugate of $U(\Lambda)$
is given by
\begin{eqnarray}
U(\Lambda)^*|\psi_0,\psi_1\rangle
&=&|\overline{v(\Lambda)_{0,0}}U(\Lambda)^*\psi_0
+\overline{v(\Lambda)_{1,0}}U(\Lambda)^*\psi_1,\cr
& &
\overline{v(\Lambda)_{0,1}}U(\Lambda)^*\psi_0
+\overline{v(\Lambda)_{1,1}}U(\Lambda)^*\psi_1\rangle
\end{eqnarray}
Note also that the representation $U(a,\Lambda)$ of $X$
is now defined by
\begin{equation}
U(a,\Lambda)=U(a,\Io)U(0,\Lambda)\equiv U(a)U(\Lambda)
\end{equation}

\subsection{Discussion}

Let $\Lambda_3(\alpha)$ denote a rotation by the angle
$\alpha$ around the third axis. It is easy to check that $v(\Lambda_3(\alpha))$
equals $\pm\exp(i(\alpha/2)\sigma_3)$ (which sign holds depends on $\alpha$ and on
the choice of inverse $v(\Lambda)$).
Now, apply twice the same rotation. Then
\begin{equation}
v(\Lambda_3(2\alpha))=\xi(\Lambda_3(\alpha),\Lambda_3(\alpha))v(\Lambda_3(\alpha))^2
\label {rot3comp}
\end{equation}
This implies that $\xi(\Lambda_3(\pi),\Lambda_3(\pi))=-1$, independent
of the choice of $v(\Lambda)$. hence, $\xi$ does take on both values $+1$ and
$-1$.

\mathenv{
Indeed, (\ref{rot3comp}) gives
\begin{eqnarray}
\Io&=&v(\Io)=v(\Lambda_3(\pi)\Lambda_3(\pi))\cr
&=&\xi(\Lambda_3(\pi),\Lambda_3(\pi))v(\Lambda_3(\pi))^2\cr
&=&\xi(\Lambda_3(\pi),\Lambda_3(\pi))e^{i\pi\sigma_3}\cr
&=&-\xi(\Lambda_3(\pi),\Lambda_3(\pi))
\end{eqnarray}

}

Under the rotation $\Lambda_3(\alpha)$ the spinor $|\psi_0,\psi_1\rangle$ transforms
into
\begin{equation}
U(\Lambda_3(\alpha))^*|\psi_0,\psi_1\rangle
=\pm |e^{-i\alpha /2}U(\Lambda_3(\alpha))\psi_0,e^{i\alpha/2}U(\Lambda_3(\alpha))\psi_1\rangle
\end{equation}
The expectation value
\begin{equation}
\langle f\rangle=\langle\psi_0|\hat f|\psi_0\rangle +
\langle\psi_1|\hat f|\psi_1\rangle
\label{val1}
\end{equation}
is mapped into
\begin{equation}
\langle U(\Lambda_3(\alpha))\psi_0|\hat f|U(\Lambda_3(\alpha))\psi_0\rangle +
\langle U(\Lambda_3(\alpha))\psi_1|\hat f|U(\Lambda_3(\alpha))\psi_1\rangle
\label{val2}
\end{equation}
and does not show any sign of the fact that a projective representation
is involved.

However, correlation functions do show a difference.
To see this, calculate
\begin{eqnarray}
& &F(f,(a,\Lambda_3(\alpha)),(b,\Lambda_3(\beta)))\cr
&=&\exp(-i(\alpha-\beta)/2)\langle\, U(a,\Lambda_3(\beta))^*\psi_0|
\hat f|U(a,\Lambda_3(\alpha))^*\psi_0\rangle\cr
&+&\exp(i(\alpha-\beta)/2)\langle\, U(a,\Lambda_3(\beta))^*\psi_1|
\hat f|U(a,\Lambda_3(\alpha))^*\psi_1\rangle
\label{spincorrel}
\end{eqnarray}
The phase factors $\exp(\pm i(\alpha-\beta)/2)$ are due to
the projective nature of the representation.

More generally one has
\begin{eqnarray}
& &F(f,(a,\Lambda),(b,M))\cr
&=&\sum_{j,j',j''}v(M)_{j',j}
\langle U(b,M)^*\psi_{j'}|\hat f|U(a,\Lambda)^*\psi_{j''}\rangle
v(\Lambda)^*_{j,j'}\cr
&=&\Tr v(\Lambda)^*Xv(M)\cr
& &\hbox{with } X_{j,j'}=
\langle U(b,M)^*\psi_{j'}|\hat f|U(a,\Lambda)^*\psi_{j}\rangle
\end{eqnarray}

\subsection{Final remarks}

Note that what is studied here is the mechanical spin of the 
electron. The magnetic moment of the electron will be discussed 
later on in the context of relativistic quantum mechanics.

The spinor representation seems to break the rotational symmetry of 
space because it favours the third direction. Indeed, the components 
$\psi_0$ and $\psi_1$ of the spinor $|\psi_0,\psi_1\rangle$ correspond 
with spin up or down in this third direction. However, this
is only apparently so. E.g., a rotation by $-\pi/2$ around
the second axis makes the first axis into the preferred direction.
Indeed, let $\Lambda=\Lambda_2(-\pi/2)$. Then one has
$v(\Lambda)=\pm 2^{-1/2} (1-i\sigma_2)$ so that
\begin{equation}
U(\Lambda)^*|\psi_0,\psi_1\rangle
=\pm 2^{-1/2}|U(\Lambda)^*(\psi_0+\psi_1),-U(\Lambda)^*(\psi_1-\psi_0)\rangle
\end{equation}
Now, the meaning of a spinor of the form $2^{-1/2}|\psi,\psi\rangle$
is that of a particle with spin up in the first direction.
After rotation it becomes $|U(\Lambda)^*\psi,0\rangle$.

The spinor representation can be reduced by restricting
the Hilbert space to spinors of the form
\begin{equation}
|\lambda\phi,\mu\phi\rangle
\end{equation}
with $\lambda$ and $\mu$ complex numbers
and $\phi$ a square integrable function. Extra symmetries must 
be considered to make it irreducible.
These are not discussed here.
The spinor representation is considered again in 
the context of relativistic quantum mechanics. There the
question of irreducibility is discussed in detail.


\section{Time evolution}

In nonrelativistic quantum mechanics time is considered as a parameter, 
not as a coordinate on the same footing as the position coordinates. As 
a consequence, we do not consider functions $f$ of position and time, 
but only of position. In other words, we reuse the classical algebra 
of functions of position and the standard representation of
quantum mechanics introduced before.

\subsection{Postulate of time evolution}

One can consider time evolution as a symmetry operation
which transforms physical states into physical states. In line with
the treatment of the group of shifts and rotations one does
also require that the representation of the classical algebra of functions
is covariant for the group of time translations. This leads to the following

\begin{postulate}[Postulate of time evolution]
\label{timeaxiom}
The time evolution is determined by unitary operators $V(t)$ which transform
any wavefunction $\psi$ into the wavefunction $\psi_t$ at time $t$ by
\begin{equation}
\psi_t=V(t)^*\psi
\label{evolvpsi}
\end{equation}
The wavefunction $\psi_t$ depends on time in a continuous manner.
The unitary operators satisfy the obvious requirements that $V(0)=\Io$
and $V(s)V(t)=V(t+s)$.
\end{postulate}

\mathenv{
For convenience, we require here that the unitary operators $V(t)$ form a
group. In principle, also projective representations should be allowed here.
}

The {\sl generator}
\index{generator of time evolution}
of the group of unitary operators $V(t)$ is denoted $\Omega$.
It is the {\sl frequency operator}.
\index{frequency operator}
It is selfadjoint: $\Omega^*=\Omega$.
By definition is
\begin{equation}
V(t)=e^{i\Omega t}
\label{evolgen}
\end{equation}
It is tradition to introduce the {\sl Hamilton operator} $H$.
\index{Hamilton operator}
It is obtained from the frequency operator by multiplication with
Planck's constant
\begin{equation}
H=\hbar\Omega
\label{hamiltonian}
\end{equation}

\mathenv{
To be more precise, the postulate of time evolution requires that
$\{V(t):\,t\in\Ro\}$ is
a strongly continuous one-parameter group of unitary operators.
The existence of $\Omega$ follows then from Stone's theorem.

} 

\subsection{Schr\"odinger equation}

From (\ref{evolvpsi}) and (\ref{evolgen}) follows immediately
\begin{equation}
i{{\rm d}\,\over{\rm d}t}\psi_t=\Omega \psi_t
\label{evoleq}
\end{equation}

\mathenv{
In many interesting cases $\Omega$ is an unbounded operator
defined on a dense domain of the Hilbert space. From Stone's
theorem follows that the domain of $\Omega$ consists
precisely of those wavefunctions $\psi_t$ for which
the derivative ${\rm d}\psi_t/{\rm d}t$ is a
quadratically integrable function.

} 

After multiplication with $\hbar$ (\ref{evoleq}) reads
\begin{equation}
i\hbar{{\rm d}\,\over{\rm d}t}\psi_t=H \psi_t
\label{Schroedeq}
\end{equation}
This is the famous Schr\"odinger equation.
Many examples involving specific choices of Hamilton operator
$H$ are discussed in the next chapter. In the next
section of the present chapter a specific expression for
$\Omega$ / $H$ describing free particle motion
is derived.

\subsection{The Heisenberg picture}

In the so called \ind{Heisenberg picture of quantum mechanics}
time evolution is associated with operators instead of with
wavefunctions. By postulate the time evolution of
any operator $A$ is given by
\begin{equation}
A_t=V(t)AV(t)^*
\end{equation}
with $V(t)$ the group of unitary operators discussed above.
In this picture , an operator $A$ without index $t$ is by convention
an operator at $t=0$; Time-dependent
wavefunctions do not occur. The relation between
the two pictures is such that for any operator $A$
and for any wavefunction $\psi$ one has
\begin{equation}
A\psi_t=(A_t\psi)_t
\label{optimevol}
\end{equation}

\mathenv{
This follows from
\begin{eqnarray}
A\psi_t&=&AV(t)^*\psi\cr
&=&V(t)^*[V(t)AV(t)^*]\psi\cr
&=&V(t)^*A_t\psi\cr
&=&(A_t\psi)_t
\end{eqnarray}
}

In particular, let $\langle\cdot\rangle_t$ denote
the quantum expectation in the state described by the
wavefunction $\psi_t$. Then one has for any operator $A$
\begin{eqnarray}
\langle A\rangle_t&=&\langle\psi_t|A|\psi_t\rangle\cr
&=&\langle\psi|A_t|\psi\cr
&=&\langle A_t\rangle
\end{eqnarray}

It is quite easy to derive a differential equation
for time-dependent operators. Taking the
time derivative of (\ref{optimevol}) one obtains
immediately
\begin{equation}
i\frac{{\rm d}\,}{{\rm d}t}A_t=[A_t,\Omega]
\end{equation}
It is tradition to multiply this equation with
Planck's constant. Then one obtains
\begin{equation}
i\hbar\frac{{\rm d}\,}{{\rm d}t}A_t=[A_t,H]
\end{equation}
This is Heisenberg's equation of motion.

\subsection{Discussion}

The two pictures, that of Schr\"odinger and that of Heisenberg,
are of course equivalent. In practical situations they are
often complementary. The formal solution of the Schr\"odinger
equation is given by (\ref{evolvpsi}). Usually
the generator $\Omega$ (or $H$) is given. Spectral theory is
then needed to calculate $V(t)$. From a numerical point of view
this can be a hopeless task. A more modest goal is then to solve
the Schr\"odinger equation for a given initial condition $\psi$
or to solve the Heisenberg equation for an initial condition $A$.
Quite often one can obtain in this way partial information about
the time evolution of a quantum particle.


\section{Mass of the particle}
\label{sectmass}

It is possible to derive the time evolution of a free quantum particle
from the principle that canonical transformations should be described
by unitary transformations (see postulate 2).
This is the topic of the present section.

\subsection{Galilei transformations}

A quantum particle must obey the same laws of physics wether described 
in a frame at rest or in a frame moving with velocity $v$. The group $X$ 
to be considered is that of the Galilei transformations, 
i.e.~general coordinate transformations described by 10 parameters 
$(a,\Lambda,t,v)$, 3 displacement components, 3 rotation angles, time, 
and 3 components of velocity.
\index{galilei group}
Its group law is
\begin{equation}
(b,M,s,w)(a,\Lambda,t,v)=
(b+M a+tw,M\Lambda,s+t,w+M v)
\end{equation}
The formula for the inverse is found to be
\begin{equation}
(a,\Lambda,t,v)^{-1}=
(-\Lambda^{-1}(a-tv),\Lambda^{-1},-t,-\Lambda^{-1}v)
\end{equation}

Can we extend the standard representation of quantum mechanics
into a representation which is covariant with respect to the full Galilei
group $X$? The surprising answer is that this is only possible
in a physically meaningfull way with
a projective representation of $X$. As discussed before, this means that there
exist complex numbers $\xi(x,y)$ of modulus one such that
\begin{equation}
U(x)U(y)=\xi(x,y)U(xy)
\end{equation}
($\xi$ is related to $\zeta$ introduced in (\ref{projective}) by
$\xi(x,y)U(xy)=U(xy)\zeta(x,y)$ --- since $\xi$ is a scalar this distinction
between left and write multiplier is not important here).
The reason why a projective representation of $X$ is acceptable is that
the physical state is only determined by the ray to which a wavefunction
$\psi$ belongs. Hence $U(xy)\psi$ and $\xi(x,y)U(xy)\psi$ refer
to the same physical state.

For the Galilei group one obtains
\begin{eqnarray}
& &U(b,M,s,w)U(a,\Lambda,t,v)\cr
&=&\xi(b,M,s,w;a,\Lambda,t,v)
U(b+M a+tw,M\Lambda,s+t,w+M v)\cr
& &
\label{galproj}
\end{eqnarray}
Only when $\xi$ is identically equal to 1 is $U$ a real representation
of the Galilei group.

\subsection{Construction of projective representations}

For each velocity $v$ let $R(v)$ denote the unitary operator
implementing the transformation to a frame moving with velocity $v$.
Let
\begin{equation}
U(0,{\Io},0,v)=R(v)
\end{equation}
and
\begin{equation}
U(0,{\Io},t,0)=V(t)
\end{equation}
and
\begin{equation}
U(a,\Lambda,0,0)=U(a,\Lambda)
\end{equation}
with $U(a,\Lambda)$ and $V(t)$ as defined in previous sections.
The projective representation is necessarily of the form
\begin{equation}
U(a,\Lambda,t,v)
=\lambda(a,\Lambda,t,v)U(a-tv,\Lambda)R(\Lambda^{-1}v)V(t)
\label{UGalDef}
\end{equation}
with $\lambda(a,\Lambda,t,v)$ a complex number of modulus 1.

\mathenv{
Indeed, using twice (\ref{galproj}), one obtains
\begin{eqnarray}
U(a,\Lambda)R(v)V(t)
&=&\xi(a,\Lambda,0,0;0,\Io,0,v)\xi(a,\Lambda,0,\Lambda v;0,\Io,t,0)\cr
&\times&
U(a+t\Lambda v,\Lambda,t,\Lambda v)
\end{eqnarray}
This implies (\ref{UGalDef}).

} 

Without restriction one can assume that $\lambda(a,\Lambda,t,v)\equiv 1$.
\mathenv{
Indeed, one can always define new operators $U'(a,\Lambda,t,v)$ by
\begin{equation}
U'(a,\Lambda,t,v)=\overline{\lambda(a,\Lambda,t,v)}U(a,\Lambda,t,v)
\end{equation}
Then $U'$ is again a projective representation, this time satisfying $\lambda\equiv 1$.
The projective representations $U$ and $U'$,
together with the representation $\hat f$ of classical functions $f$,
determine equivalent covariant representations.

} 

An appropriate choice for the unitary representation $R$ is
\begin{equation}
R(v)=e^{i\kappa (v/c)\cdot Q}
\end{equation}
with $\kappa$ some real constant, and $c$ the speed of light.
\mathenv{
Speed of light is used here to give $\kappa$ the dimension of
a wavevector. It is of course not very natural to
introduce $c$ in a non-relativistic context.
But it makes notations more consistent throughout the book.

}
A long calculation (reproduced in the appendix) shows then that
a necessary condition for 
$U(a,\Lambda,t,v)$ to be a projective representation is
that $\kappa\not=0$ and that
the generator $\Omega$ of the time evolution $V(t)$
is of the form
\begin{equation}
\Omega=\frac{c}{2\kappa}K^2 + d
\label{freemotionomega}
\end{equation}
with $d$ a real constant.
The resulting expression for $\xi$ is
\begin{equation}
\xi(b,M,s,w;a,\Lambda,t,v)=
\exp(i(\kappa/c)[w\cdot Ma -s|v|^2/2-(s+t)w\cdot Mv])
\end{equation}

\subsection{Position of the moving particle}

In the previous subsection has been shown that the standard representation
of quantum mechanics contains a projective
representation of the 10-parameter Galilei group provided that
the generator $\Omega$ of the time evolution is of the form
(\ref{freemotionomega}).

Assume now that the particle is described in the frame of the lab
at time $t=0$ by a wave function
$\psi$ which belongs to the domain of the position operators $Q_j$,
i.e.
\begin{equation}
||Q_j\psi||^2=
\int_\Ro^3\hbox{ d}q\,|q_j\psi(q)|^2<+\infty
\end{equation}
for $j=1,2,3$.
The quantum expectation value of its position at $t=0$ equals
\begin{equation}
\langle Q_j\rangle=\int_\Ro^3\hbox{ d}q\,q_j|\psi(q)|^2
\end{equation}

Consider a Galilei frame moving with speed $v$
with respect to the frame of the lab. Assume the two frames have
parallel axes and coincide at $t=0$.
The expected position in the moving frame is denoted
$\langle Q\rangle_{v,t}$ and is related to that in the frame of the
lab by
\begin{equation}
\langle Q\rangle_{v,t}=\langle Q\rangle_{0,t}-tv
\label{posmovfram}
\end{equation}

\mathenv{
The particle as observed in this moving frame at time $t$ has
wave function $U(0,\Io,t,v)^*\psi$. The observed position has components
\begin{eqnarray}
\langle Q_j\rangle_{v,t}
&=&\langle U(0,\Io,t,v)^*\psi|Q_j|U(0,\Io,t,v)^*\psi\rangle\cr
&=&\int_\Ro^3\hbox{ d}q\,q_j|U(0,\Io,t,v)^*\psi(q)|^2
\label{galobspos}
\end{eqnarray}
Now calculate
\begin{eqnarray}
U(0,\Io,t,v)^*\psi(q)
&=&V(t)^*R(v)^*U(-tv)^*\psi(q)\cr
&=&\int_{\Ro^3}\hbox{ d}k\,\tilde\phi(k)e^{ik\cdot q}e^{-i(|k|^2c/2\kappa+d)t}
\label{galtemp1}
\end{eqnarray}
with
\begin{eqnarray}
\phi(q)&=&R(v)^*U(-tv)^*\psi(q)\cr
&=&e^{-i(\kappa/c) v\cdot q}\psi(q-tv)
\end{eqnarray}
Fourier transform of the latter gives
\begin{eqnarray}
\tilde\phi(k)
&=&(2\pi)^{-3}\int_{\Ro^3}\hbox{ d}q\,e^{-ik\cdot q}\phi(q)\cr
&=&(2\pi)^{-3}\int_{\Ro^3}\hbox{ d}q\,e^{-ik\cdot q}
e^{-i(\kappa (v/c)\cdot q}\psi(q-tv)\cr
&=&(2\pi)^{-3}\int_{\Ro^3}\hbox{ d}q\,e^{-ik\cdot (q+tv)}
e^{-i\kappa (v/c)\cdot (q+tv)}\psi(q)\cr
&=&e^{-itk\cdot v}e^{-i(\kappa/c) t|v|^2}\tilde\psi(k+\kappa v/c)
\end{eqnarray}
Hence (\ref{galtemp1}) becomes
\begin{eqnarray}
& &U(0,\Io,t,v)^*\psi(q)\cr
&=&\int_{\Ro^3}\hbox{ d}k\,
e^{-itk\cdot v}e^{-i(\kappa/c) t|v|^2}
e^{ik\cdot q}e^{-i(|k|^2c/2\kappa+d)t}
\tilde\psi(k+\kappa v/c)\cr
&=&\int_{\Ro^3}\hbox{ d}k\,
e^{-it(k-\kappa v/c)\cdot v}e^{-i(\kappa/c) t|v|^2}
e^{i(k-\kappa v/c)\cdot q}\cr
& &\times e^{-i(|k-\kappa v/c|^2c/2\kappa+d)t}
\tilde\psi(k)\cr
&=&e^{-i\kappa (v/c)\cdot q}e^{-i(\kappa/c) t|v|^2/2}e^{-idt}\cr
& &\times
\int_{\Ro^3}\hbox{ d}k\,
e^{ik\cdot(q+tv)}e^{-it|k|^2c/2\kappa}
\tilde\psi(k)
\end{eqnarray}
Insertion in (\ref{galobspos}) gives
\begin{eqnarray}
\langle Q_j\rangle_{v,t}
&=&\int_\Ro^3\hbox{ d}q\,q_j
\left|
\int_{\Ro^3}\hbox{ d}k\,
e^{ik\cdot(q+tv)}e^{-it|k|^2c/2\kappa}
\tilde\psi(k)
\right|^2\cr
&=&\int_\Ro^3\hbox{ d}q\,(q-tv)_j
\left|
\int_{\Ro^3}\hbox{ d}k\,
e^{ik\cdot q}e^{-it|k|^2c/2\kappa}
\tilde\psi(k)
\right|^2\cr
&=&\langle Q_j\rangle_{0,t}-tv_j
\end{eqnarray}

} 

The previous calculation shows also that the time-dependent position of the
particle has two contributions, one of which is given by (\ref{posmovfram})
and which is the trivial
effect of transforming to a frame moving with velocity $v$,
and one which is the intrinsic evolution of the wavefunction,
given by
\begin{equation}
\langle Q_j\rangle_t=\int_\Ro^3\hbox{ d}q\,q_j
\left|
\int_{\Ro^3}\hbox{ d}k\,
e^{ik\cdot q}e^{-it|k|^2c/2\kappa}
\tilde\psi(k)
\right|^2
\end{equation}
Note that the constant $d$ in (\ref{freemotionomega}) is clearly not relevant.
It does not appear in the time evolution of the physical state
of the particle.

Let us now make a rather formal calculation in order to analyze
the behaviour of the contribution $\langle Q_j\rangle_{0,t}$.
From the CCR (see (\ref{kqccr})) follows that
\begin{equation}
e^{-ik\cdot Q}\Omega e^{ik\cdot Q}={c\over 2\kappa}(K+k)^2+d
\end{equation}
This implies
\begin{equation}
e^{-ik\cdot Q}e^{i\Omega t} e^{ik\cdot Q}=\exp(it({c\over2\kappa}(K+k)^2+d))
\end{equation}
and hence
\begin{eqnarray}
\langle e^{ik\cdot Q}\rangle_{0,t}
&=&\langle e^{-i\Omega t}\psi|e^{ik\cdot Q}|e^{-i\Omega t}\psi\rangle\cr
&=&\langle\psi|e^{ik\cdot Q}e^{it({c\over2\kappa}(K+k)^2+d)}|
e^{-i\Omega t}\psi\rangle\cr
&=&e^{it|k|^2c/2\kappa}\langle\psi|e^{ik\cdot Q}e^{it k\cdot K c/\kappa}
\psi\rangle
\end{eqnarray}
Development for small $k$ gives
\begin{equation}
\langle Q_j\rangle_{0,t}=\langle Q_j\rangle_{0,0}
+{ct\over\kappa}\langle K_j\rangle_{0,0}
\end{equation}
This shows that
$Kc/\kappa$
is the velocity operator of the particle, in the same way
as $Q$ is the position operator.
Using $P=\hbar K$, which is the conventional definition of
momentum in quantum mechanics, one sees that the mass $m$ of the particle
is given by
\begin{equation}
m=\frac{\hbar}{c} \kappa
\label{massdef}
\end{equation}

\subsection{Free equation of motion}

Note that Schr\"odinger's equation
(\ref{evoleq}), in case of the free particle, can be written as
\begin{equation}
i{{\rm d}\,\over{\rm d} t}\psi_t
=-{c\over 2\kappa}\sum_{j=1}^n{\partial^2\,\over\partial q_j^2}\psi_t
\label{freeevoleq}
\end{equation}
(use (\ref{kexplicit}) and (\ref{freemotionomega}) to see this --- we put $d=0$).
The formal solution of this equation, given a wavefunction $\psi$
at $t=0$, is of course
\begin{equation}
\psi_t=V(t)^*\psi
\end{equation}
A more explicit solution is obtained by taking $v$ equal to 0 in 
(\ref{galtemp1}). This gives
\begin{equation}
\psi_t(q)=
\int_{\Ro^3}\hbox{ d}k\,\tilde\psi(k)e^{ik\cdot q}e^{-i(|k|^2c/2\kappa)t}
\end{equation}
with $\tilde\psi$ the fourier transform of $\psi$, as before.
The common interpretation of this formula is that the plane wave 
\index{wavevector}
$e^{ik\cdot q}$ evolves in time by multiplication with a phase
factor $e^{-i(|k|^2c/2\kappa)t}$. To obtain the time evolution
of a wavefunction $\psi$ it suffices therefore to
decompose it into plane waves and to insert the time evolution of the latter.

\subsection{Time reversal}

Time reversal maps the element $(a,\Lambda,t,v)$ of the 10-parameter group
onto the element $(a,\Lambda,-t,-v)$. It is a symmetry of classical mechanics.
Any solution of the equations of motions of classical mechanics is again a
solution after reversal of time. It should also be a symmetry of quantum mechanics.
First we verify that under time reversal a particle with mass $\kappa>0$
transforms into a particle with mass $\kappa<0$. This is seen immediately
from the definition (see (\ref{UGalDef}) --- $\lambda$ is taken equal to 1) that
\begin{equation}
U(a,\Lambda,t,v)=U(a-tv,\Lambda)R(\Lambda^{-1}v)V(t)
\end{equation}
with
\begin{equation}
R(v)=e^{i\kappa (v/c)\cdot Q}
\qquad\hbox{ and }
V(t)=e^{i\Omega t}=e^{it|K|^2c/2\kappa}
\end{equation}
(we take the constant $d$ in (\ref{freemotionomega}) equal to zero).

Since time reversal is a symmetry of quantum mechanics there should exists
a unitary operator $\theta$ of the Hilbert space of wave functions implementing this
symmetry, at least if the posutlate (\ref{postcov}) holds. It turns out
that the postulate must be weakened slightly. Wigner\cite{WEP31} showed that
each symmetry operation must be implemented either by a unitary or by an anti-unitary
operator. In case of continuous symmetries, as considered so far, one can exclude
the anti-unitary case. However, time reversal is a discrete symmetry. It turns out
that the operator $\theta$ which we are looking for maps each wavefunction $\psi$
onto its complex conjugate $\overline\psi$. It clearly is anti-unitary.

Take the complex conjugate of (\ref{freeevoleq}). One obtains
\begin{equation}
-i{{\rm d}\,\over{\rm d} t}\overline\psi_t
=-{c\over 2\kappa}\sum_{j=1}^n{\partial^2\,\over\partial q_j^2}\overline \psi_t
\end{equation}
It is clear that $\overline \psi_{-t}$ satisfies Schr\"odinger's equation again.
Also $\overline \psi_{t}$ satisfies Schr\"odinger's equation
but this time for a particle with mass $-\kappa$. One concludes that the
anti-unitary operator $\theta$ implements time reversal.


\section{Quantum particle on a circle}

In this section we study the motion of a quantum particle on the circle.
Starting point for the description are functions $f(\alpha)$ where
$\alpha$ is an angle between 0 and $2\pi$.
A rotation by an angle $\alpha'$ transforms the function $f(\alpha)$
into the function $g(\alpha)$ given by
\begin{equation}
g(\alpha)=f(\alpha-\alpha' \,{\rm modulo}\, 2\pi)
\label{circlerot}
\end{equation}
The symmetry group $X$ is the group of rotations SO$(1)$.

The description of the particle on the circle is
very similar to that of a particle on the real line.
The main novelty that occurs is that the wavevector operator
$K$ and the frequency operator $\Omega$ (resp.~the momentum operator
$P$ and the Hamilton operator $H$) have purely discrete spectra.
This means that there exists an orthonormal basis $(\psi_n)_n$ of the Hilbert space
of wavefunctions which diagonalises $K$ and $\Omega$.
In the terminology of classical mechanics,
the quantum particle can be at rest at specific discrete
values of the energy. One says that the energy is
quantised. A quantum of energy is the amount of energy needed
to bring a particle at rest again at rest, but now at a higher energy level.
This quantisation of energy is the origin of the name of quantum mechanics.

\mathenv{
The mathematical reason for the important difference
between a particle on the real line and one on the circle
is that in the latter case the group of symmetries $X$ is compact.
Its dual group ${\rm Z},+$ is discrete.

}

\subsection{Representation}

The obvious choice of Hilbert space consists of square integrable
functions $\psi(\alpha)$ which depend on the angle $\alpha$.
A function $f(\alpha)$ is represented by the operator $\hat f$
which maps $\psi(\alpha)$ onto $f(\alpha)\psi(\alpha)$.
A unitary representation $U(\alpha)$ of SO$(1)$ is defined by
\begin {equation}
U(\alpha')\psi(\alpha)=\psi(\alpha-\alpha' \,{\rm modulo}\, 2\pi)
\end{equation}

\mathenv{
It satisfies
\begin{eqnarray}
U(\alpha')\hat fU(\alpha')^*\psi(\alpha)
&=& \hat fU(\alpha')^*\psi(\alpha-\alpha'\,{\rm modulo}\, 2\pi)\cr
&=&g(\alpha)U(\alpha')^*\psi(\alpha-\alpha'\,{\rm modulo}\, 2\pi)\cr
&=&g(\alpha)\psi(\alpha)\cr
&=&\hat g\psi(\alpha)
\end{eqnarray}
where $g(\alpha)$ is given by (\ref{circlerot}).
This shows that it implements rotation of functions $f(\alpha)$.

}

An orthonormal basis of the Hilbert space is formed by the wavefunctions
\begin{equation}
\psi_n(\alpha)=\frac{1}{\sqrt{2\pi}}e^{in\alpha}
\end{equation}
Obviously, one has
\begin{eqnarray}
U(\alpha')\psi_n(\alpha)&=&\frac{1}{\sqrt{2\pi}}e^{in(\alpha-\alpha')}\cr
&=&e^{-in\alpha'}\psi_n(\alpha)
\label{eigenualpha}
\end{eqnarray}
This shows that the wavefunctions $\psi_n$ are eigenfunctions of the operator
$U(\alpha')$.

\subsection{CCR and free particle motion}

Let $K$ denote the generator of rotations
\begin{equation}
U(\alpha)=e^{-i\alpha K}
\end{equation}
(note the minus sign in the definition of $K$).
Then (\ref{eigenualpha}) shows that 
\begin{equation}
K\psi_n(\alpha)=n\psi_n(\alpha)
\end{equation}
i.e., the operator $K$ has
a discrete spectrum with eigenvectors $\psi_n$,
and with corresponding eigenvalue the integer $n$.

The position operator $Q$ is defined by
\begin{equation}
Q\psi(\alpha)=\alpha\psi(\alpha)
\end{equation}
The operators $Q$ and $K$ satisfy the same canonical commutation relations (CCR)
as in the case of a particle on the real line.

The frequency operator for the free particle motion is
conventionally taken equal to
\begin{equation}
\Omega=\frac{c}{2\kappa} K^2
\end{equation}
with $\kappa$ a positive constant ($(\hbar/c)\kappa=m$ is 
the mass of the particle). Clearly, the
frequency operator $\Omega$ has a discrete spectrum.
The eigenvalues are $n^2c/2\kappa$ with $n$ an integer number.
Each eigenvalue, except for zero, is twofold degenerate.
The time evolution of the eigenfunctions $\psi_n$ is given by
\begin{equation}
\left(\psi_n\right)_t=e^{-ict/2\kappa}\psi_n
\end{equation}
Multiplying $\psi_n$ with the complex
phase factor $\exp(-ict/2\kappa)$ does not
change the state of the particle. Hence, the state
of the particle described by the eigenfunction $\psi_n$
does not depend on time. In other words the particle is
at rest. The correlation functions in this state
are given by
\begin{equation}
F(f,\alpha',\alpha'')=e^{in(\alpha'-\alpha'')}
\frac{1}{2\pi}\int_0^{2\pi}\hbox{ d}\alpha\,f(\alpha)
\end{equation}


\section*{Notes to Chapter \Operatorformalism}
\addcontentsline{toc}{section}{Notes to Chapter \Operatorformalism}


\paragraph{Functions of position}

Quantum mechanics is introduced here starting with functions $f(q)$ of 
position $q$. This might give the impression that only functions of 
position are experimentally measurable. This is not the point of view 
taken here. All information about the physical state of the quantum 
particle is encoded in the correlation functions $F(f,a,b)$. In 
principle, this information is experimentally accessible. Two physical 
states differ if they are described by different correlation 
functions. It is understood of course that, if two states differ, then 
it is feasible, at least in principle, to make the difference by 
experimental means. This point of view is the standard one. New is that 
we emphazise that the notion of physical state depends on the symmetry 
group $X$ under consideration. In particular, properties like spin and 
mass of the particle lable different states of the same particle. This 
requires that one uses the Galilei group as the relevant symmetry group. 
In this context a physical state is not longer uniquely determined by a 
ray in Hilbert space. We support our point of view by noting that spin 
and mass can be measured experimentally so that in our approach there is 
still a one-to-one relation between physical states which are 
experimentally distinguishable and correlation functions $F(f,x,y)$.




\paragraph{Symmetry} Symmetry is the holy grail of theoretical physics. 
Our picture of the world is idealized by introducing symmetry 
principles, eventually noting that they are not perfectly satisfied. In 
elementary particle physics Lie groups and Lie algebras are dominant 
tools. Representations of groups are used to classify particles. In the 
present formalism the group of symmetry elements is not considered on 
itself but acts on an algebra of observable quantities. The emphasis lies 
on combined representations of the algebra, together with the group 
acting on it. These are called covariant representations. Their 
importance to physics has been brought up by Doplicher et al 
\cite{DKR66}.

\paragraph{Correlation functions}
Correlation functions form the corner stone of the
mathematical approach to quantum mechanics known under the
name of $C^*$-algebraic approach. In this context they are called
\indf{mathematical states}. Each mathematical state induces
a Hilbert space representation through the
\indf{Gelfand-Naimark-Segal (GNS) theorem}.

The material presented in this book is mostly based on results obtained 
in the $C^*$-algebraic formalism. Of particular influence was the 
formulation of quantum mechanics and quantum field theory in terms of 
the Weyl algebra. See Petz \cite{PD90} for an introduction to these 
matters. In \cite {NJ99}, representations of the Weyl algebra are 
considered as covariant representations of a smaller algebra. This is 
the basic idea underlying the present approach to quantum mechanics and 
quantum field theory. In Naudts and Kuna \cite {NK00} the notion of 
mathematical state of a $C^*$-algebra is generalized to that of 
correlation functions $F(f,x,y)$ of a covariance system. An accompanying 
generalized GNS-theorem is also proved.




\paragraph{Planck's constant}
Throughout the book the abundant use of Planck's constant $\hbar$ is
avoided by dividing classical quantities like
energy, mass, and momentum, with $\hbar$. The more common alternative
is to choose units such that $\hbar=1$.
Planck's constant links the absolute units of quantum mechanics to the
units of classical mechanics which historically have been chosen
without knowing their mutual relations. In particular, 
$\hbar$ is a manifestation of the relation between mass and frequency resp.~inverse distance,
in a similar way as speed of light $c$ is a relation between time and distance.

One often says that classical mechanics is obtained from quantum mechanics
by taking the limit taking $\hbar$ equal to zero. If literally taken,
this is obviously wrong since quantum mechanics can be formulated
in such a way that $\hbar$ does not appear.
On the contrary, one must introduce $\hbar$ to introduce classical
concepts like energy, mass, and momentum.


\paragraph{Rotation symmetry}
An introductory text explaining the rotation group is
found in \cite{SW86}. The time spent on studying rotation
symmetry may seem exagerated for a chapter introducing basic
quantum mechanics. However, spin of quantum particles cannot
be understood correctly without going through rotation symmetry
with attention to sufficient detail.


\paragraph{Spin}

It is common to describe spin as an internal property
of the particle like mass or charge, and, at the sime time,
as an extra degree of freedom attached to the particle.
This is quite strange because charge or mass are
not considered to be extra degrees of freedom.
In the present book al three properties are
treated on the same footing, being labels
for different representations of a given particle.



\paragraph{Mass of the particle}
The discovery that mass is a parameter labeling projective representations
of the Galilei group goes back to Levy-Leblond \cite{LL63}.



\section*{Exercises}
\addcontentsline{toc}{section}{Exercises}
\begin{enumerate}

\item
Expression (\ref{Ftochar}) gives the correlation function
$F(f,a,b)$ in terms of the characteristic function $\chi(k,a)$.
It can be inverted, giving $\chi(k,a)$ in terms of $F(f,a,b)$.

\mathenv{
Introduce new variables $c=(a+b)/2$ and $d=b-a$. Then (\ref{Ftochar}) becomes
\begin{equation}
F(f, c-d/2,c+d/2)=
\int_{\Ro^n}\hbox{ d}k\,\tilde f(k)e^{-ik\cdot c}
\chi(k,d)
\end{equation}
Fourier transforming this expression gives
\begin{equation}
(2\pi)^n\tilde f(k)\chi(k,d)=\int\hbox{ d}c\, e^{ic\cdot k}F(f,c-d/2,c+d/2)
\end{equation}
Now, make an explicit choice of $f(q)$, e.g. $f=g$ with
\begin{equation}
g(q)=(2\pi)^{-n/2}\exp(-(1/2)|q|^2)
\end{equation}
It satisfies $\tilde g(k)=(2\pi)^{-n}\exp(-(1/2)|k|^2)$.
The desired relation reads now
\begin{equation}
\chi(k,d)=e^{(1/2)|k|^2}\int_{-\infty}^{+\infty}\hbox{ d}c\,
e^{ic\cdot k}F(g,c-d/2,c+d/2)
\end{equation}

}

\item
Show that for a particle without spin $F(f,(a,\Lambda),(b,M))$ can be
expressed in terms of correlation functions $F(f,a,b)$ which involve
only shifts in space $a$ and $b$ (see section (\ref{rotsymsection})).

\mathenv{
Calculate
\begin{eqnarray}
& &F(f,a,b)\cr
&=&\langle\psi|U(b)\hat fU(a)^*|\psi\rangle\cr
&=&\int\hbox{ d}q\,f(q)\psi(q+a)\overline{\psi(q+b)}\cr
&=&\int\hbox{ d}q\,f(q)\int\hbox{ d}k\,e^{ik\cdot (q+a)}\tilde\psi(k)
\int\hbox{ d}k'\,e^{-ik'\cdot (q+b)}
\overline{\tilde\psi(q+b)}\cr
&=&(2\pi)^n\int\hbox{ d}k\,e^{ik\cdot a}\tilde\psi(k)
\int\hbox{ d}k'\,e^{-ik'\cdot b}
\overline{\tilde\psi(k')}
\tilde f(k'-k)
\end{eqnarray}
Fourier transforming this expression w.r.t.~$a$ and $b$ gives
\begin{eqnarray}
& &\tilde\psi(k)\overline{\tilde\psi(k')}\tilde f(k'-k)\cr
&=&(2\pi)^{-n}\int\hbox{ d}a\int\hbox{ d}b\,
e^{-ik\cdot a}e^{ik'\cdot b} F(f,a,b)
\end{eqnarray}
Now make the same choice of function $f$ as in the previous
exercise. There follows
\begin{eqnarray}
\tilde\psi(k)\overline{\tilde\psi(k')}
&=&e^{(1/2)|k'-k|^2}\int\hbox{ d}a\int\hbox{ d}b\,
e^{-ik\cdot a}e^{ik'\cdot b} F(g,a,b)\cr
& & 
\label{exc1.2l1}
\end{eqnarray}

Next calculate
\begin{eqnarray}
& &F(f,(a,\Lambda),(b,M))\cr
&=&\langle\psi|U(b,M)\hat f U(a,\Lambda)^*|\psi\rangle\cr
&=&\int\hbox{ d}q\,f(q)\psi(\Lambda q+a)\overline{\psi(Mq+b)}\cr
&=&\int\hbox{ d}q\,f(q)\cr
& &\times
\int\hbox{ d}k\,\tilde\psi(k) e^{ik\cdot (\Lambda q+a)}
\int\hbox{ d}k'\,\overline{\tilde\psi(k')}e^{-ik'\cdot (Mq+b)}\cr
& &
\end{eqnarray}
In combination with (\ref{exc1.2l1}) this gives
\begin{eqnarray}
& &F(f,(a,\Lambda),(b,M))\cr
&=&\int\hbox{ d}q\,f(q)
\int\hbox{ d}k\, e^{ik\cdot (\Lambda q+a)}
\int\hbox{ d}k'\,e^{-ik'\cdot (Mq+b)}\cr
& &\times e^{(1/2)|k'-k|^2}\int\hbox{ d}a'\int\hbox{ d}b'\,
e^{-ik\cdot a'}e^{ik'\cdot b'} F(g,a',b')\cr
&=&(2\pi)^n\int\hbox{ d}k\int\hbox{ d}k'\int\hbox{ d}a'\int\hbox{ d}b'\,
e^{ik\cdot (a-a')}e^{-ik'\cdot (b-b')}\cr
& &\times e^{(1/2)|k'-k|^2}\tilde f(M^{-1}k'-\Lambda^{-1}k) F(g,a',b')
\end{eqnarray}

}

\item
Consider a free quantum particle on the real line $\Ro$.
Solve Heisenberg's equations of motion for the position and
wavevector operators $Q$ resp.~$K$.

\mathenv{
$\Omega$ equals $(c/2\kappa)K^2$. One calculates that
\begin{equation}
[Q,\Omega]=(ic/\kappa)K\quad\hbox{ and }[K,\Omega]=0
\end{equation}
Now, $\Omega$ is time-independent. Hence the equations
of motion can be written as
\begin{equation}
i\frac{{\rm d}\,}{{\rm d}t}A_t=([A,\Omega])_t
\end{equation}
In particular one obtains
\begin{eqnarray}
& &i\frac{{\rm d}\,}{{\rm d}t}Q_t=\frac{ic}{\kappa}K_t\cr
& &i\frac{{\rm d}\,}{{\rm d}t}K_t=0
\end{eqnarray}
The solution of the latter equation is $K_t=K$.
Integration of the former equation gives
\begin{equation}
Q_t=Q + (tc/\kappa)K
\end{equation}
This shows that $(c/\kappa)K$ is the velocity operator.
This was noted already in section (\ref{sectmass}).

}

\end{enumerate}

\section*{Appendix
A --- Construction of a projective representation}
\addcontentsline{toc}{section}{Appendix A}

From the definitions follows immediately that
\begin{eqnarray}
U(a,\Lambda,0,v)\psi(q)
&=&e^{i\kappa (v/c)\cdot (q-a)}\psi(\Lambda^{-1}(q-a))
\end{eqnarray}
Using this expresssion one calculates
\begin{eqnarray}
& &U(b,M,0,w)U(a,\Lambda,0,v)\psi(q)\cr
&=&e^{i\kappa (w/c)\cdot (q-b)}U(a,\Lambda,0,v)\psi((M)^{-1}(q-b))\cr
&=&e^{i\kappa (w/c)\cdot (q-b)}e^{i\kappa (v/c)\cdot ((M)^{-1}(q-b)-a)}
\psi(\Lambda^{-1}((M)^{-1}(q-b)-a))
\end{eqnarray}
and
\begin{eqnarray}
& &U(b+M a,M\Lambda,0,w+M v)\psi(q)\cr
&=&\exp(i\kappa ((w+M v)/c)\cdot(q-b-M a)))\cr
&\times&
\psi((M\Lambda)^{-1}(q-b-M a))
\end{eqnarray}
Comparing both using (\ref{galproj}) gives
\begin{eqnarray}
\xi(b,M,0,w;a,\Lambda,0,v)
&=&e^{i\kappa (w/c)\cdot (M a)}
\end{eqnarray}
One verifies easily that $\xi$ as given by this expression
is a cocycle of the 9-parameter group not including time.

Finally let us extend $\xi$ so as to contain time.
Because $V(t)$ commutes with $U(a,\Lambda)$ there exists a function
$f$ such that $\Omega=f(K)$. This implies
\begin{equation}
V(t)=e^{i\Omega t}=e^{if(K)t}
\end{equation}
Hence,
\begin{equation}
V(t)R(v)\psi(q)=\int_{\Ro^n}\hbox{ d}k\,e^{ik\cdot q}e^{if(k)t}\tilde\phi(k)
\end{equation}
with
\begin{equation}
\phi(q)=e^{i\kappa (v/c)\cdot q}\psi(q)
\end{equation}
From
\begin{eqnarray}
\tilde\phi(k)
&=&(2\pi)^{-n}\int_{\Ro^n}\hbox{ d}q\,
e^{-ik\cdot q}e^{i\kappa (v/c)\cdot q}\psi(q)\cr
&=&\tilde\psi(k-\kappa (v/c))
\end{eqnarray}
follows then
\begin{eqnarray}
V(t)R(v)\psi(q)
&=&\int_{\Ro^n}\hbox{ d}k\,e^{ikq}e^{if(k)t}\tilde\psi(k-\kappa v/c)\cr
&=&\int_{\Ro^n}\hbox{ d}k\,e^{i(k+\kappa v/c)\cdot q}e^{if(k+\kappa v/c)t}\tilde\psi(k)
\label{ap1btmp1}
\end{eqnarray}
On the other hand is
\begin{equation}
V(t)R(v)=\xi(\Io,0,t,0;\Io,0,0,v)U(-tv,\Io)R(v)V(t)
\label{ap1btmp2}
\end{equation}
and
\begin{eqnarray}
& &U(-tv,\Io)R(v)V(t)\psi(q)\cr
&=&R(v)V(t)\psi(q+tv)\cr
&=&e^{i\kappa (v/c)\cdot (q+tv)}V(t)\psi(q+tv)\cr
&=&e^{i\kappa (v/c)\cdot (q+tv)}\int_{\Ro^n}\hbox{ d}k\,
e^{ik\cdot (q+tv)}e^{if(k)t}\tilde\psi(k)
\label{ap1btmp3}
\end{eqnarray}
Because $\psi$ is arbitrary one concludes
from (\ref{ap1btmp1}), (\ref{ap1btmp2}) and (\ref{ap1btmp3}) that
\begin{equation}
e^{i(k+\kappa v/c)\cdot q}e^{if(k+\kappa v/c)t}
=\xi(0,\Io,t,0;0,\Io,0,v)
e^{i\kappa (v/c)\cdot (q+tv)}e^{ik\cdot (q+tv)}e^{if(k)t}
\end{equation}
This can be written as
\begin{equation}
\xi(0,\Io,t,0;0,\Io,0,v)
=e^{-itk\cdot v}e^{-if(k)t}e^{if(k+\kappa v/c)t}
\end{equation}
Note that $\xi$ does not depend on $k$.
This gives the set of equations
(assuming that $t\not=0$)
\begin{equation}
{\partial f\over\partial k_j}(k+\kappa v/c)
={\partial f\over\partial k_j}(k)+v_j
\end{equation}
The solution is a quadratic form
\begin{equation}
f(k)={c\over 2\kappa}|k|^2+\beta \cdot k +d
\end{equation}
No solution exists if $\kappa=0$ and $v\not=0$.

Let us now calculate $\xi(b,M,s,w;a,\Lambda,t,v)$.
Start with evaluation of
\begin{eqnarray}
& &U(a,\Lambda,t,v)\psi(q)\cr
&=&U(a-tv,\Lambda)R(\Lambda^{-1}v)V(t)\psi(q)\cr
&=&e^{i\kappa (v/c)\cdot (q-a+tv)}
\int_{\Ro^n}\hbox{ d}k\,\tilde \psi(k)\cr
& &\times
\exp(ik\cdot \Lambda^{-1}(q-a+tv))
\exp(it(|k|^2c/2\kappa+\beta \cdot k+d))\cr
&=&
\int_{\Ro^n}\hbox{ d}k\,\tilde \psi(\Lambda^{-1}(k-\kappa v/c))
\exp(ik\cdot (q-a+tv))\cr
& &\times
\exp(it(|k-\kappa v|^2c/2\kappa+\Lambda \beta \cdot (k-\kappa v/c)+d))
\end{eqnarray}
Let $\psi'(q)=U(a,\Lambda,t,v)\psi(q)$.
One obtains from the previous expression
\begin{eqnarray}
\tilde\psi'(k)&=&
e^{-ik\cdot (a-tv)}\tilde\psi(\Lambda^{-1}(k-\kappa v/c))\cr
& &\times
\exp(it(|k-\kappa v/c|^2c/2\kappa+(\Lambda \beta )\cdot(k-\kappa v/c)+d)))
\end{eqnarray}
Using this result twice one obtains
\begin{eqnarray}
U(b,M,s,w)U(a,\Lambda,t,v)\psi(q)
&=&U(b,M,s,w)\psi'(q)\cr
&=&\psi''(q)
\end{eqnarray}
with
\begin{eqnarray}
\tilde\psi''(k)
&=&\tilde\psi'(M^{-1}(k-\kappa w/c))e^{-ik\cdot (b-sw)}\cr
&\times&
\exp(is(|k-\kappa w/c|^2c/2\kappa +M\beta \cdot (k-\kappa w/c)+d))\cr
&=&e^{-iM^{-1}(k-\kappa w/c)\cdot (a-tv)}
\tilde\psi(\Lambda^{-1}(M^{-1}(k-\kappa w/c)-\kappa v/c))\cr
&\times& e^{it(|M^{-1}(k-\kappa w/c)-\kappa v/c|^2c/2\kappa
+\Lambda \beta \cdot(M^{-1}(k-\kappa w/c)-\kappa v/c)+d)}\cr
&\times&e^{-ik\cdot (b-sw)}
e^{is(|k-\kappa w/c|^2c/2\kappa +M\beta \cdot (k-\kappa w/c)+d)}
\label{coc1}
\end{eqnarray}
On the other hand is, using (\ref{galproj}),
\begin{eqnarray}
& &U(b,M,s,w)U(a,\Lambda,t,v)\psi(q)\cr
&=&\xi(b,M,s,w;a,\Lambda,t,v)
U(b+Ma+tw,M\Lambda,s+t,w+Mv)\psi(q)\cr
&=&\xi(b,M,s,w;a,\Lambda,t,v)\psi'''(q)
\end{eqnarray}
with
\begin{eqnarray}
\tilde\psi'''(k)&=&
e^{-ik\cdot (b+Ma+tw-(s+t)(w+Mv))}\tilde\psi((M\Lambda)^{-1}(k-\kappa (w+Mv)/c))\cr
& &\times
e^{i(s+t)(|k-\kappa (w+Mv)/c|^2c/2\kappa+(M\Lambda \beta)\cdot(k-\kappa (w+Mv)/c)+d))}
\end{eqnarray}
The equality $\psi''(q)=\xi(b,M,s,w;a,\Lambda,t,v)\psi'''(q)$ should hold.
This yields an expression for $\xi$ which should not depend on $k$.
This is only possible if $\beta=0$. Then one obtains
\begin{eqnarray}
& &\xi(b,M,s,w;a,\Lambda,t,v)\cr
&=&\exp(i\kappa[w\cdot Ma -s|v|^2/2c-(s+t)w\cdot Mv])
\end{eqnarray}


\end{document}